\documentclass[11pt]{article}
\include{thesis.preamble}
\usepackage{dsfont}
\usepackage{yfonts}
\usepackage{rotating}
\usepackage{floatpag}
\rotfloatpagestyle{empty}

\usepackage{hyperref}

\usepackage{amsmath,amsthm,amssymb}
\usepackage{xspace,enumerate,color,epsfig}
\usepackage{graphicx}
\graphicspath{{.}{./figures/}}
\usepackage{marvosym}

\usepackage{tikzfig}
\usepackage{stmaryrd}
\usepackage{docmute}
\usepackage{keycommand}

\usepackage{pifont}

\usepackage{enumitem}

\newkeycommand{\pointmap}[style=point][1]{\,\tikz{\node[style=\commandkey{style}] (x) at (0,-0.3) {$#1$};\node [style=none] (3) at (0, 1.0) {};\node [style=none] (3) at (0, -1.0) {};\draw (x)--(0,0.7);}\,}

\newkeycommand{\typedkpoint}[style=kpoint,edgestyle=][2]{%
\,\begin{tikzpicture}
  \begin{pgfonlayer}{nodelayer}
    \node [style=none] (0) at (0, 1) {};
    \node [style=\commandkey{style}] (1) at (0, -0.75) {$#2$};
    \node [style=right label] (2) at (0.25, 0.5) {$#1$};
  \end{pgfonlayer}
  \begin{pgfonlayer}{edgelayer}
    \draw [\commandkey{edgestyle}] (1) to (0.center);
  \end{pgfonlayer}
\end{tikzpicture}\,}

\newkeycommand{\typedkpointdag}[style=kpoint adjoint,edgestyle=][2]{%
\,\begin{tikzpicture}
  \begin{pgfonlayer}{nodelayer}
    \node [style=none] (0) at (0, -1) {};
    \node [style=\commandkey{style}] (1) at (0, 0.75) {$#2$};
    \node [style=right label] (2) at (0.25, -0.5) {$#1$};
  \end{pgfonlayer}
  \begin{pgfonlayer}{edgelayer}
    \draw [\commandkey{edgestyle}] (0.center) to (1);
  \end{pgfonlayer}
\end{tikzpicture}\,}

\newcommand{\bistateadj}[1]{\,\begin{tikzpicture}[yshift=-3mm]
  \begin{pgfonlayer}{nodelayer}
    \node [style=kpointadj, minimum width=1 cm, inner sep=2pt] (0) at (0, 0.5) {$\psi$};
    \node [style=none] (1) at (-0.75, 0.25) {};
    \node [style=none] (2) at (0.75, 0.25) {};
    \node [style=none] (3) at (-0.75, -0.5) {};
    \node [style=none] (4) at (0.75, -0.5) {};
  \end{pgfonlayer}
  \begin{pgfonlayer}{edgelayer}
    \draw (1.center) to (3.center);
    \draw (2.center) to (4.center);
  \end{pgfonlayer}
\end{tikzpicture}\,}

\newkeycommand{\pointketbra}[style1=copoint,style2=point][2]{\,%
\begin{tikzpicture}
  \begin{pgfonlayer}{nodelayer}
    \node [style=none] (0) at (0, -1.5) {};
    \node [style=\commandkey{style1}] (1) at (0, -0.75) {$#1$};
    \node [style=\commandkey{style2}] (2) at (0, 0.75) {$#2$};
    \node [style=none] (3) at (0, 1.5) {};
  \end{pgfonlayer}
  \begin{pgfonlayer}{edgelayer}
    \draw (0.center) to (1);
    \draw (2) to (3.center);
  \end{pgfonlayer}
\end{tikzpicture}\,}

\newkeycommand{\twocopointketbra}[style1=copoint,style2=copoint,style3=point][3]{\,%
\begin{tikzpicture}
  \begin{pgfonlayer}{nodelayer}
    \node [style=none] (0) at (-0.75, -1.5) {};
    \node [style=none] (1) at (0.75, -1.5) {};
    \node [style=\commandkey{style1}] (2) at (-0.75, -0.75) {$#1$};
    \node [style=\commandkey{style2}] (3) at (0.75, -0.75) {$#2$};
    \node [style=\commandkey{style3}] (4) at (0, 0.75) {$#3$};
    \node [style=none] (5) at (0, 1.5) {};
  \end{pgfonlayer}
  \begin{pgfonlayer}{edgelayer}
    \draw (0.center) to (2);
    \draw (1.center) to (3);
    \draw (4) to (5.center);
  \end{pgfonlayer}
\end{tikzpicture}\,}

\newkeycommand{\twopointketbra}[style1=copoint,style2=point,style3=point][3]{\,%
\begin{tikzpicture}
  \begin{pgfonlayer}{nodelayer}
    \node [style=none] (0) at (0, -1.5) {};
    \node [style=none] (1) at (-0.75, 1.5) {};
    \node [style=\commandkey{style1}] (2) at (0, -0.75) {$#1$};
    \node [style=\commandkey{style2}] (3) at (-0.75, 0.75) {$#2$};
    \node [style=\commandkey{style3}] (4) at (0.75, 0.75) {$#3$};
    \node [style=none] (5) at (0.75, 1.5) {};
  \end{pgfonlayer}
  \begin{pgfonlayer}{edgelayer}
    \draw (0.center) to (2);
    \draw (1.center) to (3);
    \draw (4) to (5.center);
  \end{pgfonlayer}
\end{tikzpicture}\,}

\newkeycommand{\pointbraket}[style1=point,style2=copoint][2]{\,%
\begin{tikzpicture}
  \begin{pgfonlayer}{nodelayer}
    \node [style=\commandkey{style1}] (0) at (0, -0.5) {$#1$};
    \node [style=\commandkey{style2}] (1) at (0, 0.5) {$#2$};
  \end{pgfonlayer}
  \begin{pgfonlayer}{edgelayer}
    \draw (0) to (1);
  \end{pgfonlayer}
\end{tikzpicture}\,}

\newkeycommand{\kpointbraket}[style1=kpoint,style2=kpoint adjoint][2]{\,%
\begin{tikzpicture}
  \begin{pgfonlayer}{nodelayer}
    \node [style=\commandkey{style1}] (0) at (0, -0.75) {$#1$};
    \node [style=\commandkey{style2}] (1) at (0, 0.75) {$#2$};
  \end{pgfonlayer}
  \begin{pgfonlayer}{edgelayer}
    \draw (0) to (1);
  \end{pgfonlayer}
\end{tikzpicture}\,}

\newkeycommand{\kpointmap}[style1=kpoint,style2=map][2]{\,%
\begin{tikzpicture}
  \begin{pgfonlayer}{nodelayer}
    \node [style=\commandkey{style1}] (0) at (0, -1.1) {$#1$};
    \node [style=\commandkey{style2}] (1) at (0, 0.2) {$#2$};
    \node [style=none] (2) at (0, 1.5) {};
  \end{pgfonlayer}
  \begin{pgfonlayer}{edgelayer}
    \draw (0) to (1);
    \draw (1) to (2.center);
  \end{pgfonlayer}
\end{tikzpicture}%
\,}

\newkeycommand{\sandwichtwo}[style1=point,style2=map,style3=map,style4=copoint][4]{\,%
\begin{tikzpicture}
  \begin{pgfonlayer}{nodelayer}
    \node [style=\commandkey{style2}] (0) at (0, -0.75) {$#2$};
    \node [style=\commandkey{style3}] (1) at (0, 0.75) {$#3$};
    \node [style=\commandkey{style1}] (2) at (0, -2) {$#1$};
    \node [style=\commandkey{style4}] (3) at (0, 2) {$#4$};
  \end{pgfonlayer}
  \begin{pgfonlayer}{edgelayer}
    \draw (2) to (0);
    \draw (0) to (1);
    \draw (1) to (3);
  \end{pgfonlayer}
\end{tikzpicture}\,}

\newkeycommand{\longbraket}[style1=point,style2=copoint][2]{\,%
\begin{tikzpicture}
  \begin{pgfonlayer}{nodelayer}
    \node [style=\commandkey{style1}] (0) at (0, -2) {$#1$};
    \node [style=\commandkey{style2}] (1) at (0, 2) {$#2$};
  \end{pgfonlayer}
  \begin{pgfonlayer}{edgelayer}
    \draw (0) to (1);
  \end{pgfonlayer}
\end{tikzpicture}\,}

\newkeycommand{\onb}[style=point]{\ensuremath{\left\{\tikz{\node[style=\commandkey{style}] (x) at (0,-0.3) {$j$};\draw (x)--(0,0.7);}\right\}}\xspace}

\newkeycommand{\copointmap}[style=copoint][1]{\,\tikz{\node[style=\commandkey{style}] (x) at (0,0.3) {$#1$};\draw (x)--(0,-0.7);}\,}

\newcommand{\bra}[1]{\ensuremath{\left\langle #1 \right|}}
\newcommand{\ket}[1]{\ensuremath{\left|  #1 \right\rangle}}

\newcommand{\ketbra}[2]{\ensuremath{\ket{#1}\!\bra{#2}}}

\tikzstyle{cdiag}=[matrix of math nodes, row sep=3em, column sep=3em, text height=1.5ex, text depth=0.25ex,inner sep=0.5em]
\tikzstyle{arrow above}=[transform canvas={yshift=0.5ex}]
\tikzstyle{arrow below}=[transform canvas={yshift=-0.5ex}]

\usetikzlibrary{shapes.multipart}

\tikzstyle{arrowhead}=[regular polygon,regular polygon sides=3,draw,scale=0.2,inner sep=-0.15pt,minimum width=6mm,fill=black,regular polygon rotate=180]
\tikzstyle{trace}=[circuit ee IEC,thick,ground,rotate=0,scale=2]
\tikzstyle{wavy}=[decorate,decoration={snake, segment length=1mm, amplitude=0.3mm}]
\tikzstyle{mopoint}=[shape=semicircle, fill=white,draw=black,shape border rotate=180,scale =0.75]
\tikzstyle{mocopoint}=[shape=semicircle, fill=white,draw=black,minimum width = 0.9cm, scale =0.75, xscale=0.7]
\tikzstyle{cpoint}=[shape=semicircle, fill=white,draw=black,minimum width = 0.9cm, scale =0.75, xscale=1, yscale=0.7, shape border rotate = 90,font=\fontsize{14}{16}\selectfont]
\tikzstyle{cocpoint}=[shape=semicircle, fill=white,draw=black,minimum width = 0.9cm, scale =0.75, xscale=1, yscale=0.7, shape border rotate = 270,font=\fontsize{14}{16}\selectfont]
\tikzstyle{slit}=[line width=2]
\tikzstyle{block}=[line width=4,red,line cap=round]
\tikzstyle{screen}=[line width=4,black,line cap=round]
\tikzstyle{di}=[diamond,draw,inner sep=0.5pt,font=\small, minimum size = .5cm]
\tikzstyle{sbox}=[rectangle,draw]
\tikzstyle{mirror}=[line width=2,black]
\tikzstyle{traceState}=[circuit ee IEC,thick,ground,rotate=180,scale=2]
\tikzstyle{detEff}=[circuit ee IEC,thick,ground,rotate=180,scale=1.4]
\tikzstyle{maxMix}=[circuit ee IEC,thick,ground,scale=1.4]
\tikzstyle{particlePath}=[line width=2,gray!40, line cap =round]

\tikzstyle{bwSpider}=[
       rectangle split,
       rectangle split parts=2,
       rectangle split part fill={black,white},
 minimum size=3.6 mm, inner sep=-2mm, draw=black,scale=0.5,rounded corners=0.8 mm
       ]
 \tikzstyle{wbSpider}=[
       rectangle split,
       rectangle split parts=2,
       rectangle split part fill={white,black},
 minimum size=3.6 mm, inner sep=-2mm, draw=black,scale=0.5,rounded corners=0.8 mm
       ]
\tikzstyle{cWire}=[densely dotted, thick]
\tikzstyle{env}=[copoint,regular polygon rotate=0,minimum width=0.2cm, fill=black]

\tikzstyle{probs}=[shape=semicircle,fill=white,draw=black,shape border rotate=180,minimum width=1.2cm]

\tikzstyle{every picture}=[baseline=-0.25em,scale=0.5]
\tikzstyle{dotpic}=[] 
\tikzstyle{diredges}=[every to/.style={diredge}]
\tikzstyle{math matrix}=[matrix of math nodes,left delimiter=(,right delimiter=),inner sep=2pt,column sep=1em,row sep=0.5em,nodes={inner sep=0pt},text height=1.5ex, text depth=0.25ex]

\tikzstyle{inline text}=[text height=1.5ex, text depth=0.25ex,yshift=0.5mm]
\tikzstyle{label}=[font=\footnotesize,text height=1.5ex, text depth=0.25ex,yshift=0.5mm]
\tikzstyle{left label}=[label,anchor=east,xshift=1.mm]
\tikzstyle{right label}=[label,anchor=west,xshift=-1.mm]

\tikzstyle{braceedge}=[decorate,decoration={brace,amplitude=2mm,raise=-1mm}]
\tikzstyle{small braceedge}=[decorate,decoration={brace,amplitude=1mm,raise=-1mm}]

\tikzstyle{doubled}=[line width=1.6pt] 
\tikzstyle{boldedge}=[doubled,shorten <=-0.17mm,shorten >=-0.17mm]
\tikzstyle{boldedgegray}=[doubled,gray,shorten <=-0.17mm,shorten >=-0.17mm]
\tikzstyle{singleedgegray}=[gray]

\tikzstyle{semidoubled}=[line width=1.4pt] 
\tikzstyle{semiboldedgegray}=[semidoubled,gray,shorten <=-0.17mm,shorten >=-0.17mm]

\tikzstyle{boxedge}=[semiboldedgegray]

\tikzstyle{boldedgedashed}=[very thick,dashed,shorten <=-0.17mm,shorten >=-0.17mm]
\tikzstyle{vboldedgedashed}=[doubled,dashed,shorten <=-0.17mm,shorten >=-0.17mm]
\tikzstyle{left hook arrow}=[left hook-latex]
\tikzstyle{right hook arrow}=[right hook-latex]
\tikzstyle{sembracket}=[line width=0.5pt,shorten <=-0.07mm,shorten >=-0.07mm]

\tikzstyle{causal edge}=[->,thick,gray]
\tikzstyle{causal nondir}=[thick,gray]
\tikzstyle{timeline}=[thick,gray, dashed]

\tikzstyle{cedge}=[<->,thick,gray!70!white]

\tikzstyle{empty diagram}=[draw=gray!40!white,dashed,shape=rectangle,minimum width=1cm,minimum height=1cm]
\tikzstyle{empty diagram small}=[draw=gray!50!white,dashed,shape=rectangle,minimum width=0.6cm,minimum height=0.5cm]

\tikzstyle{dot}=[inner sep=0mm,minimum width=2mm,minimum height=2mm,draw,shape=circle]

\tikzstyle{leak}=[white dot, shape=regular polygon, minimum size=3.3 mm, regular polygon sides=3, outer sep=-0.2mm, regular polygon rotate=270]
\tikzstyle{proj}=[white dot, shape=regular polygon, minimum size=3.3 mm, regular polygon sides=4, outer sep=-0.2mm]
\tikzstyle{Vleak}=[white dot, shape=regular polygon, minimum size=3.3 mm, regular polygon sides=3, outer sep=-0.2mm, regular polygon rotate=90]
\tikzstyle{dleak}=[white dot, line width=1.6pt, shape=regular polygon, minimum size=3.3 mm, regular polygon sides=3, outer sep=-0.2mm, regular polygon rotate=270]

\tikzstyle{Wsquare}=[white dot, shape=regular polygon, rounded corners=0.8 mm, minimum size=3.3 mm, regular polygon sides=3, outer sep=-0.2mm]
\tikzstyle{Wsquareadj}=[white dot, shape=regular polygon, rounded corners=0.8 mm, minimum size=3.3 mm, regular polygon sides=3, outer sep=-0.2mm, regular polygon rotate=180]
\tikzstyle{ddot}=[inner sep=0mm, doubled, minimum width=2.5mm,minimum height=2.5mm,draw,shape=circle]

\tikzstyle{black dot}=[dot,fill=black]
\tikzstyle{white dot}=[dot,fill=white,,text depth=-0.2mm]
\tikzstyle{white Wsquare}=[Wsquare,fill=gray,,text depth=-0.2mm]
\tikzstyle{white Wsquareadj}=[Wsquareadj,fill=white,,text depth=-0.2mm]
\tikzstyle{green dot}=[white dot] 
\tikzstyle{gray dot}=[dot,fill=gray!40!white,,text depth=-0.2mm]
\tikzstyle{red dot}=[gray dot] 

\tikzstyle{black ddot}=[ddot,fill=black]
\tikzstyle{white ddot}=[ddot,fill=white]
\tikzstyle{gray ddot}=[ddot,fill=gray!40!white]

\tikzstyle{gray edge}=[gray!60!white]

\tikzstyle{small dot}=[inner sep=0.5mm,minimum width=0pt,minimum height=0pt,draw,shape=circle]

\tikzstyle{small black dot}=[small dot,fill=black]
\tikzstyle{small white dot}=[small dot,fill=white]
\tikzstyle{small gray dot}=[small dot,fill=gray!40!white]

\tikzstyle{causal dot}=[inner sep=0.4mm,minimum width=0pt,minimum height=0pt,draw=white,shape=circle,fill=gray!40!white]

\tikzstyle{phase dimensions}=[minimum size=5mm,font=\footnotesize,rectangle,rounded corners=2.5mm,inner sep=0.2mm,outer sep=-2mm]
\tikzstyle{dphase dimensions}=[minimum size=5mm,font=\footnotesize,rectangle,rounded corners=2.5mm,inner sep=0.2mm,outer sep=-2mm]

\tikzstyle{white phase dot}=[dot,fill=white,phase dimensions]
\tikzstyle{white phase ddot}=[ddot,fill=white,dphase dimensions]

\tikzstyle{white rect ddot}=[draw=black,fill=white,doubled,minimum size=5mm,font=\footnotesize,rectangle,rounded corners=2.5mm,inner sep=0.2mm]
\tikzstyle{gray rect ddot}=[draw=black,fill=gray!40!white,doubled,minimum size=6mm,font=\footnotesize,rectangle,rounded corners=3mm]

\tikzstyle{gray phase dot}=[dot,fill=gray!40!white,phase dimensions]
\tikzstyle{gray phase ddot}=[ddot,fill=gray!40!white,dphase dimensions]
\tikzstyle{grey phase dot}=[gray phase dot]
\tikzstyle{grey phase ddot}=[gray phase ddot]

\tikzstyle{small phase dimensions}=[minimum size=4mm,font=\tiny,rectangle,rounded corners=2mm,inner sep=0.2mm,outer sep=-2mm]
\tikzstyle{small dphase dimensions}=[minimum size=4mm,font=\tiny,rectangle,rounded corners=2mm,inner sep=0.2mm,outer sep=-2mm]

\tikzstyle{small gray phase dot}=[dot,fill=gray!40!white,small phase dimensions]
\tikzstyle{small gray phase ddot}=[ddot,fill=gray!40!white,small dphase dimensions]

\tikzstyle{small map}=[draw,shape=rectangle,minimum height=4mm,minimum width=4mm,fill=white]

\tikzstyle{cnot}=[fill=white,shape=circle,inner sep=-1.4pt]

\tikzstyle{asym hadamard}=[fill=white,draw,shape=NEbox,inner sep=0.6mm,font=\footnotesize,minimum height=4mm]
\tikzstyle{asym hadamard conj}=[fill=white,draw,shape=NWbox,inner sep=0.6mm,font=\footnotesize,minimum height=4mm]
\tikzstyle{asym hadamard dag}=[fill=white,draw,shape=SEbox,inner sep=0.6mm,font=\footnotesize,minimum height=4mm]

\tikzstyle{hadamard}=[fill=white,draw,inner sep=0.6mm,font=\footnotesize,minimum height=4mm,minimum width=4mm]
\tikzstyle{small hadamard}=[fill=white,draw,inner sep=0.6mm,minimum height=1.5mm,minimum width=1.5mm]
\tikzstyle{small hadamard rotate}=[small hadamard,rotate=45]
\tikzstyle{dhadamard}=[hadamard,doubled]
\tikzstyle{small dhadamard}=[small hadamard,doubled]
\tikzstyle{small dhadamard rotate}=[small hadamard rotate,doubled]
\tikzstyle{antipode}=[white dot,inner sep=0.3mm,font=\footnotesize]

\tikzstyle{scalar}=[diamond,draw,inner sep=0.5pt,font=\small]
\tikzstyle{dscalar}=[diamond,doubled, draw,inner sep=0.5pt,font=\small]

\tikzstyle{small box}=[rectangle,inline text,fill=white,draw,minimum height=5mm,yshift=-0.5mm,minimum width=5mm,font=\small]
\tikzstyle{small gray box}=[small box,fill=gray!30]
\tikzstyle{medium box}=[rectangle,inline text,fill=white,draw,minimum height=5mm,yshift=-0.5mm,minimum width=10mm,font=\small]
\tikzstyle{square box}=[small box] 
\tikzstyle{medium gray box}=[small box,fill=gray!30]
\tikzstyle{semilarge box}=[rectangle,inline text,fill=white,draw,minimum height=5mm,yshift=-0.5mm,minimum width=12.5mm,font=\small]
\tikzstyle{large box}=[rectangle,inline text,fill=white,draw,minimum height=5mm,yshift=-0.5mm,minimum width=15mm,font=\small]
\tikzstyle{large gray box}=[small box,fill=gray!30]

\tikzstyle{Bayes box}=[rectangle,fill=black,draw, minimum height=3mm, minimum width=3mm]

\tikzstyle{gray square point}=[small box,fill=gray!50]

\tikzstyle{dphase box white}=[dhadamard]
\tikzstyle{dphase box gray}=[dhadamard,fill=gray!50!white]
\tikzstyle{phase box white}=[hadamard]
\tikzstyle{phase box gray}=[hadamard,fill=gray!50!white]

\tikzstyle{point}=[regular polygon,regular polygon sides=3,draw,scale=0.75,inner sep=-0.5pt,minimum width=9mm,fill=white,regular polygon rotate=180]
\tikzstyle{point nosep}=[regular polygon,regular polygon sides=3,draw,scale=0.75,inner sep=-2pt,minimum width=9mm,fill=white,regular polygon rotate=180]
\tikzstyle{copoint}=[regular polygon,regular polygon sides=3,draw,scale=0.75,inner sep=-0.5pt,minimum width=9mm,fill=white]
\tikzstyle{dpoint}=[point,doubled]
\tikzstyle{dcopoint}=[copoint,doubled]

\tikzstyle{pointgrow}=[shape=cornerpoint,kpoint common,scale=0.75,inner sep=3pt]
\tikzstyle{pointgrow dag}=[shape=cornercopoint,kpoint common,scale=0.75,inner sep=3pt]

\tikzstyle{wide copoint}=[fill=white,draw,shape=isosceles triangle,shape border rotate=90,isosceles triangle stretches=true,inner sep=0pt,minimum width=1.5cm,minimum height=6.12mm]
\tikzstyle{wide point}=[fill=white,draw,shape=isosceles triangle,shape border rotate=-90,isosceles triangle stretches=true,inner sep=0pt,minimum width=1.5cm,minimum height=6.12mm,yshift=-0.0mm]
\tikzstyle{wide point plus}=[fill=white,draw,shape=isosceles triangle,shape border rotate=-90,isosceles triangle stretches=true,inner sep=0pt,minimum width=1.74cm,minimum height=7mm,yshift=-0.0mm]

\tikzstyle{wide dpoint}=[fill=white,doubled,draw,shape=isosceles triangle,shape border rotate=-90,isosceles triangle stretches=true,inner sep=0pt,minimum width=1.5cm,minimum height=6.12mm,yshift=-0.0mm]

\tikzstyle{tinypoint}=[regular polygon,regular polygon sides=3,draw,scale=0.55,inner sep=-0.15pt,minimum width=6mm,fill=white,regular polygon rotate=180]

\tikzstyle{white point}=[point]
\tikzstyle{white dpoint}=[dpoint]
\tikzstyle{green point}=[white point] 
\tikzstyle{white copoint}=[copoint]
\tikzstyle{gray point}=[point,fill=gray!40!white]
\tikzstyle{gray dpoint}=[gray point,doubled]
\tikzstyle{red point}=[gray point] 
\tikzstyle{gray copoint}=[copoint,fill=gray!40!white]
\tikzstyle{gray dcopoint}=[gray copoint,doubled]

\tikzstyle{white point guide}=[regular polygon,regular polygon sides=3,font=\scriptsize,draw,scale=0.65,inner sep=-0.5pt,minimum width=9mm,fill=white,regular polygon rotate=180]

\tikzstyle{black point}=[point,fill=black,font=\color{white}]
\tikzstyle{black copoint}=[copoint,fill=black,font=\color{white}]

\tikzstyle{tiny gray point}=[tinypoint,fill=gray!40!white]

\tikzstyle{diredge}=[->]
\tikzstyle{ddiredge}=[<->]
\tikzstyle{rdiredge}=[<-]
\tikzstyle{thickdiredge}=[->, very thick]
\tikzstyle{pointer edge}=[->,very thick,gray]
\tikzstyle{pointer edge part}=[very thick,gray]
\tikzstyle{dashed edge}=[dashed]
\tikzstyle{thick dashed edge}=[very thick,dashed]
\tikzstyle{thick gray dashed edge}=[thick dashed edge,gray!40]
\tikzstyle{thick map edge}=[very thick,|->]

\makeatletter
\newcommand{\boxshape}[3]{%
\pgfdeclareshape{#1}{
\inheritsavedanchors[from=rectangle] 
\inheritanchorborder[from=rectangle]
\inheritanchor[from=rectangle]{center}
\inheritanchor[from=rectangle]{north}
\inheritanchor[from=rectangle]{south}
\inheritanchor[from=rectangle]{west}
\inheritanchor[from=rectangle]{east}
\backgroundpath{
\southwest \pgf@xa=\pgf@x \pgf@ya=\pgf@y
\northeast \pgf@xb=\pgf@x \pgf@yb=\pgf@y

\@tempdima=#2
\@tempdimb=#3

\pgfpathmoveto{\pgfpoint{\pgf@xa - 5pt + \@tempdima}{\pgf@ya}}
\pgfpathlineto{\pgfpoint{\pgf@xa - 5pt - \@tempdima}{\pgf@yb}}
\pgfpathlineto{\pgfpoint{\pgf@xb + 5pt + \@tempdimb}{\pgf@yb}}
\pgfpathlineto{\pgfpoint{\pgf@xb + 5pt - \@tempdimb}{\pgf@ya}}
\pgfpathlineto{\pgfpoint{\pgf@xa - 5pt + \@tempdima}{\pgf@ya}}
\pgfpathclose
}
}}

\boxshape{NEbox}{0pt}{5pt}
\boxshape{SEbox}{0pt}{-5pt}
\boxshape{NWbox}{5pt}{0pt}
\boxshape{SWbox}{-5pt}{0pt}
\boxshape{EBox}{-3pt}{3pt}
\boxshape{WBox}{3pt}{-3pt}
\makeatother

\tikzstyle{cloud}=[shape=cloud,draw,minimum width=1.5cm,minimum height=1.5cm]

\tikzstyle{map}=[draw,shape=NEbox,inner sep=2pt,minimum height=6mm,fill=white]
\tikzstyle{dashedmap}=[draw,dashed,shape=NEbox,inner sep=2pt,minimum height=6mm,fill=white]
\tikzstyle{mapdag}=[draw,shape=SEbox,inner sep=2pt,minimum height=6mm,fill=white]
\tikzstyle{mapadj}=[draw,shape=SEbox,inner sep=2pt,minimum height=6mm,fill=white]
\tikzstyle{maptrans}=[draw,shape=SWbox,inner sep=2pt,minimum height=6mm,fill=white]
\tikzstyle{mapconj}=[draw,shape=NWbox,inner sep=2pt,minimum height=6mm,fill=white]

\tikzstyle{medium map}=[draw,shape=NEbox,inner sep=2pt,minimum height=6mm,fill=white,minimum width=7mm]
\tikzstyle{medium map dag}=[draw,shape=SEbox,inner sep=2pt,minimum height=6mm,fill=white,minimum width=7mm]
\tikzstyle{medium map adj}=[draw,shape=SEbox,inner sep=2pt,minimum height=6mm,fill=white,minimum width=7mm]
\tikzstyle{medium map trans}=[draw,shape=SWbox,inner sep=2pt,minimum height=6mm,fill=white,minimum width=7mm]
\tikzstyle{medium map conj}=[draw,shape=NWbox,inner sep=2pt,minimum height=6mm,fill=white,minimum width=7mm]
\tikzstyle{semilarge map}=[draw,shape=NEbox,inner sep=2pt,minimum height=6mm,fill=white,minimum width=9.5mm]
\tikzstyle{semilarge map trans}=[draw,shape=SWbox,inner sep=2pt,minimum height=6mm,fill=white,minimum width=9.5mm]
\tikzstyle{semilarge map adj}=[draw,shape=SEbox,inner sep=2pt,minimum height=6mm,fill=white,minimum width=9.5mm]
\tikzstyle{semilarge map dag}=[draw,shape=SEbox,inner sep=2pt,minimum height=6mm,fill=white,minimum width=9.5mm]
\tikzstyle{semilarge map conj}=[draw,shape=NWbox,inner sep=2pt,minimum height=6mm,fill=white,minimum width=9.5mm]
\tikzstyle{large map}=[draw,shape=NEbox,inner sep=2pt,minimum height=6mm,fill=white,minimum width=12mm]
\tikzstyle{large map conj}=[draw,shape=NWbox,inner sep=2pt,minimum height=6mm,fill=white,minimum width=12mm]
\tikzstyle{very large map}=[draw,shape=NEbox,inner sep=2pt,minimum height=6mm,fill=white,minimum width=17mm]

\tikzstyle{medium dmap}=[draw,doubled,shape=NEbox,inner sep=2pt,minimum height=6mm,fill=white,minimum width=7mm]
\tikzstyle{medium dmap dag}=[draw,doubled,shape=SEbox,inner sep=2pt,minimum height=6mm,fill=white,minimum width=7mm]
\tikzstyle{medium dmap adj}=[draw,doubled,shape=SEbox,inner sep=2pt,minimum height=6mm,fill=white,minimum width=7mm]
\tikzstyle{medium dmap trans}=[draw,doubled,shape=SWbox,inner sep=2pt,minimum height=6mm,fill=white,minimum width=7mm]
\tikzstyle{medium dmap conj}=[draw,doubled,shape=NWbox,inner sep=2pt,minimum height=6mm,fill=white,minimum width=7mm]
\tikzstyle{semilarge dmap}=[draw,doubled,shape=NEbox,inner sep=2pt,minimum height=6mm,fill=white,minimum width=9.5mm]
\tikzstyle{semilarge dmap trans}=[draw,doubled,shape=SWbox,inner sep=2pt,minimum height=6mm,fill=white,minimum width=9.5mm]
\tikzstyle{semilarge dmap adj}=[draw,doubled,shape=SEbox,inner sep=2pt,minimum height=6mm,fill=white,minimum width=9.5mm]
\tikzstyle{semilarge dmap dag}=[draw,doubled,shape=SEbox,inner sep=2pt,minimum height=6mm,fill=white,minimum width=9.5mm]
\tikzstyle{semilarge dmap conj}=[draw,doubled,shape=NWbox,inner sep=2pt,minimum height=6mm,fill=white,minimum width=9.5mm]
\tikzstyle{large dmap}=[draw,doubled,shape=NEbox,inner sep=2pt,minimum height=6mm,fill=white,minimum width=12mm]
\tikzstyle{large dmap conj}=[draw,doubled,shape=NWbox,inner sep=2pt,minimum height=6mm,fill=white,minimum width=12mm]
\tikzstyle{large dmap trans}=[draw,doubled,shape=SWbox,inner sep=2pt,minimum height=6mm,fill=white,minimum width=12mm]
\tikzstyle{large dmap adj}=[draw,doubled,shape=SEbox,inner sep=2pt,minimum height=6mm,fill=white,minimum width=12mm]
\tikzstyle{large dmap dag}=[draw,doubled,shape=SEbox,inner sep=2pt,minimum height=6mm,fill=white,minimum width=12mm]
\tikzstyle{very large dmap}=[draw,doubled,shape=NEbox,inner sep=2pt,minimum height=6mm,fill=white,minimum width=19.5mm]

\tikzstyle{muxbox}=[draw,shape=rectangle,minimum height=3mm,minimum width=3mm,fill=white]
\tikzstyle{dmuxbox}=[muxbox,doubled]

\tikzstyle{box}=[draw,shape=rectangle,inner sep=2pt,minimum height=6mm,minimum width=6mm,fill=white]
\tikzstyle{dbox}=[draw,doubled,shape=rectangle,inner sep=2pt,minimum height=6mm,minimum width=6mm,fill=white]
\tikzstyle{dmap}=[draw,doubled,shape=NEbox,inner sep=2pt,minimum height=6mm,fill=white]
\tikzstyle{dmapdag}=[draw,doubled,shape=SEbox,inner sep=2pt,minimum height=6mm,fill=white]
\tikzstyle{dmapadj}=[draw,doubled,shape=SEbox,inner sep=2pt,minimum height=6mm,fill=white]
\tikzstyle{dmaptrans}=[draw,doubled,shape=SWbox,inner sep=2pt,minimum height=6mm,fill=white]
\tikzstyle{dmapconj}=[draw,doubled,shape=NWbox,inner sep=2pt,minimum height=6mm,fill=white]

\tikzstyle{ddmap}=[draw,doubled,dashed,shape=NEbox,inner sep=2pt,minimum height=6mm,fill=white]
\tikzstyle{ddmapdag}=[draw,doubled,dashed,shape=SEbox,inner sep=2pt,minimum height=6mm,fill=white]
\tikzstyle{ddmapadj}=[draw,doubled,dashed,shape=SEbox,inner sep=2pt,minimum height=6mm,fill=white]
\tikzstyle{ddmaptrans}=[draw,doubled,dashed,shape=SWbox,inner sep=2pt,minimum height=6mm,fill=white]
\tikzstyle{ddmapconj}=[draw,doubled,dashed,shape=NWbox,inner sep=2pt,minimum height=6mm,fill=white]

\boxshape{sNEbox}{0pt}{3pt}
\boxshape{sSEbox}{0pt}{-3pt}
\boxshape{sNWbox}{3pt}{0pt}
\boxshape{sSWbox}{-3pt}{0pt}
\tikzstyle{smap}=[draw,shape=sNEbox,fill=white]
\tikzstyle{smapdag}=[draw,shape=sSEbox,fill=white]
\tikzstyle{smapadj}=[draw,shape=sSEbox,fill=white]
\tikzstyle{smaptrans}=[draw,shape=sSWbox,fill=white]
\tikzstyle{smapconj}=[draw,shape=sNWbox,fill=white]

\tikzstyle{dsmap}=[draw,dashed,shape=sNEbox,fill=white]
\tikzstyle{dsmapdag}=[draw,dashed,shape=sSEbox,fill=white]
\tikzstyle{dsmaptrans}=[draw,dashed,shape=sSWbox,fill=white]
\tikzstyle{dsmapconj}=[draw,dashed,shape=sNWbox,fill=white]

\boxshape{mNEbox}{0pt}{10pt}
\boxshape{mSEbox}{0pt}{-10pt}
\boxshape{mNWbox}{10pt}{0pt}
\boxshape{mSWbox}{-10pt}{0pt}
\tikzstyle{mmap}=[draw,shape=mNEbox]
\tikzstyle{mmapdag}=[draw,shape=mSEbox]
\tikzstyle{mmaptrans}=[draw,shape=mSWbox]
\tikzstyle{mmapconj}=[draw,shape=mNWbox]

\tikzstyle{mmapgray}=[draw,fill=gray!40!white,shape=mNEbox]
\tikzstyle{smapgray}=[draw,fill=gray!40!white,shape=sNEbox]

\makeatletter

\pgfdeclareshape{cornerpoint}{
\inheritsavedanchors[from=rectangle] 
\inheritanchorborder[from=rectangle]
\inheritanchor[from=rectangle]{center}
\inheritanchor[from=rectangle]{north}
\inheritanchor[from=rectangle]{south}
\inheritanchor[from=rectangle]{west}
\inheritanchor[from=rectangle]{east}
\backgroundpath{
\southwest \pgf@xa=\pgf@x \pgf@ya=\pgf@y
\northeast \pgf@xb=\pgf@x \pgf@yb=\pgf@y

\pgfmathsetmacro{\pgf@shorten@left}{\pgfkeysvalueof{/tikz/shorten left}}
\pgfmathsetmacro{\pgf@shorten@right}{\pgfkeysvalueof{/tikz/shorten right}}

\pgfpathmoveto{\pgfpoint{0.5 * (\pgf@xa + \pgf@xb)}{\pgf@ya - 5pt}}
\pgfpathlineto{\pgfpoint{\pgf@xa - 8pt + \pgf@shorten@left}{\pgf@yb - 1.5 * \pgf@shorten@left}}
\pgfpathlineto{\pgfpoint{\pgf@xa - 8pt + \pgf@shorten@left}{\pgf@yb}}
\pgfpathlineto{\pgfpoint{\pgf@xb + 8pt - \pgf@shorten@right}{\pgf@yb}}
\pgfpathlineto{\pgfpoint{\pgf@xb + 8pt - \pgf@shorten@right}{\pgf@yb - 1.5 * \pgf@shorten@right}}
\pgfpathclose
}
}

\pgfdeclareshape{cornercopoint}{
\inheritsavedanchors[from=rectangle] 
\inheritanchorborder[from=rectangle]
\inheritanchor[from=rectangle]{center}
\inheritanchor[from=rectangle]{north}
\inheritanchor[from=rectangle]{south}
\inheritanchor[from=rectangle]{west}
\inheritanchor[from=rectangle]{east}
\backgroundpath{
\southwest \pgf@xa=\pgf@x \pgf@ya=\pgf@y
\northeast \pgf@xb=\pgf@x \pgf@yb=\pgf@y

\pgfmathsetmacro{\pgf@shorten@left}{\pgfkeysvalueof{/tikz/shorten left}}
\pgfmathsetmacro{\pgf@shorten@right}{\pgfkeysvalueof{/tikz/shorten right}}

\pgfpathmoveto{\pgfpoint{0.5 * (\pgf@xa + \pgf@xb)}{\pgf@yb + 5pt}}
\pgfpathlineto{\pgfpoint{\pgf@xa - 8pt + \pgf@shorten@left}{\pgf@ya + 1.5 * \pgf@shorten@left}}
\pgfpathlineto{\pgfpoint{\pgf@xa - 8pt + \pgf@shorten@left}{\pgf@ya}}
\pgfpathlineto{\pgfpoint{\pgf@xb + 8pt - \pgf@shorten@right}{\pgf@ya}}
\pgfpathlineto{\pgfpoint{\pgf@xb + 8pt - \pgf@shorten@right}{\pgf@ya + 1.5 * \pgf@shorten@right}}
\pgfpathclose
}
}

\makeatother

\pgfkeyssetvalue{/tikz/shorten left}{0pt}
\pgfkeyssetvalue{/tikz/shorten right}{0pt}

\tikzstyle{kpoint common}=[draw,fill=white,inner sep=1pt,minimum height=4mm]
\tikzstyle{kpoint sc}=[shape=cornerpoint,kpoint common]
\tikzstyle{kpoint adjoint sc}=[shape=cornercopoint,kpoint common]
\tikzstyle{kpoint}=[shape=cornerpoint,shorten left=5pt,kpoint common]
\tikzstyle{kpoint adjoint}=[shape=cornercopoint,shorten left=5pt,kpoint common]
\tikzstyle{kpoint conjugate}=[shape=cornerpoint,shorten right=5pt,kpoint common]
\tikzstyle{kpoint transpose}=[shape=cornercopoint,shorten right=5pt,kpoint common]
\tikzstyle{kpoint symm}=[shape=cornerpoint,shorten left=5pt,shorten right=5pt,kpoint common]

\tikzstyle{wide kpoint sc}=[shape=cornerpoint,kpoint common, minimum width=1 cm]
\tikzstyle{wide kpointdag sc}=[shape=cornercopoint,kpoint common, minimum width=1 cm]

\tikzstyle{black kpoint}=[shape=cornerpoint,shorten left=5pt,kpoint common,fill=black,font=\color{white}]

\tikzstyle{black kpoint sm}=[shape=cornerpoint,shorten left=5pt,kpoint common,fill=black,font=\color{white},scale=0.75]

\tikzstyle{black kpoint adjoint}=[shape=cornercopoint,shorten left=5pt,kpoint common,fill=black,font=\color{white}]
\tikzstyle{black kpointadj}=[shape=cornercopoint,shorten left=5pt,kpoint common,fill=black,font=\color{white}]

\tikzstyle{black kpointadj sm}=[shape=cornercopoint,shorten left=5pt,kpoint common,fill=black,font=\color{white},scale=0.75]

\tikzstyle{black dkpoint}=[shape=cornerpoint,shorten left=5pt,kpoint common,fill=black, doubled,font=\color{white}]
\tikzstyle{black dkpoint adjoint}=[shape=cornercopoint,shorten left=5pt,kpoint common,fill=black, doubled,font=\color{white}]
\tikzstyle{black dkpointadj}=[shape=cornercopoint,shorten left=5pt,kpoint common,fill=black, doubled,font=\color{white}]

\tikzstyle{black dkpoint sm}=[shape=cornerpoint,shorten left=5pt,kpoint common,fill=black, doubled,font=\color{white},scale=0.75]
\tikzstyle{black dkpointadj sm}=[shape=cornercopoint,shorten left=5pt,kpoint common,fill=black, doubled,font=\color{white},scale=0.75]

\tikzstyle{kpointdag}=[kpoint adjoint]
\tikzstyle{kpointadj}=[kpoint adjoint]
\tikzstyle{kpointconj}=[kpoint conjugate]
\tikzstyle{kpointtrans}=[kpoint transpose]

\tikzstyle{big kpoint}=[kpoint, minimum width=1.2 cm, minimum height=8mm, inner sep=4pt, text depth=3mm]

\tikzstyle{wide kpoint}=[kpoint, minimum width=1 cm, inner sep=2pt]
\tikzstyle{wide kpointdag}=[kpointdag, minimum width=1 cm, inner sep=2pt]
\tikzstyle{wide kpointconj}=[kpointconj, minimum width=1 cm, inner sep=2pt]
\tikzstyle{wide kpointtrans}=[kpointtrans, minimum width=1 cm, inner sep=2pt]

\tikzstyle{wider kpoint}=[kpoint, minimum width=1.25 cm, inner sep=2pt]
\tikzstyle{wider kpointdag}=[kpointdag, minimum width=1.25 cm, inner sep=2pt]
\tikzstyle{wider kpointconj}=[kpointconj, minimum width=1.25 cm, inner sep=2pt]
\tikzstyle{wider kpointtrans}=[kpointtrans, minimum width=1.25 cm, inner sep=2pt]

\tikzstyle{gray kpoint}=[kpoint,fill=gray!50!white]
\tikzstyle{gray kpointdag}=[kpointdag,fill=gray!50!white]
\tikzstyle{gray kpointadj}=[kpointadj,fill=gray!50!white]
\tikzstyle{gray kpointconj}=[kpointconj,fill=gray!50!white]
\tikzstyle{gray kpointtrans}=[kpointtrans,fill=gray!50!white]

\tikzstyle{gray dkpoint}=[kpoint,fill=gray!50!white,doubled]
\tikzstyle{gray dkpointdag}=[kpointdag,fill=gray!50!white,doubled]
\tikzstyle{gray dkpointadj}=[kpointadj,fill=gray!50!white,doubled]
\tikzstyle{gray dkpointconj}=[kpointconj,fill=gray!50!white,doubled]
\tikzstyle{gray dkpointtrans}=[kpointtrans,fill=gray!50!white,doubled]

\tikzstyle{white label}=[draw,fill=white,rectangle,inner sep=0.7 mm]
\tikzstyle{gray label}=[draw,fill=gray!50!white,rectangle,inner sep=0.7 mm]
\tikzstyle{black label}=[draw,fill=black,rectangle,inner sep=0.7 mm]

\tikzstyle{dkpoint}=[kpoint,doubled]
\tikzstyle{wide dkpoint}=[wide kpoint,doubled]
\tikzstyle{dkpointdag}=[kpoint adjoint,doubled]
\tikzstyle{wide dkpointdag}=[wide kpointdag,doubled]
\tikzstyle{dkcopoint}=[kpoint adjoint,doubled]
\tikzstyle{dkpointadj}=[kpoint adjoint,doubled]
\tikzstyle{dkpointconj}=[kpoint conjugate,doubled]
\tikzstyle{dkpointtrans}=[kpoint transpose,doubled]

\tikzstyle{kscalar}=[kpoint common, shape=EBox, inner xsep=-1pt, inner ysep=3pt,font=\small]
\tikzstyle{kscalarconj}=[kpoint common, shape=WBox, inner xsep=-1pt, inner ysep=3pt,font=\small]

\tikzstyle{spekpoint}=[kpoint sc,minimum height=5mm,inner sep=3pt]
\tikzstyle{spekcopoint}=[kpoint adjoint sc,minimum height=5mm,inner sep=3pt]

\tikzstyle{dspekpoint}=[spekpoint,doubled]
\tikzstyle{dspekcopoint}=[spekcopoint,doubled]

 \tikzstyle{upground}=[circuit ee IEC,thick,ground,rotate=90,scale=2.5]
 \tikzstyle{downground}=[circuit ee IEC,thick,ground,rotate=-90,scale=2.5]
 \tikzstyle{bigground}=[regular polygon,regular polygon sides=3,draw=gray,scale=0.50,inner sep=-0.5pt,minimum width=10mm,fill=gray]

\tikzstyle{arrs}=[-latex,font=\small,auto]
\tikzstyle{arrow plain}=[arrs]
\tikzstyle{arrow dashed}=[dashed,arrs]
\tikzstyle{arrow bold}=[very thick,arrs]
\tikzstyle{arrow hide}=[draw=white!0,-]
\tikzstyle{arrow reverse}=[latex-]
\tikzstyle{cdnode}=[]

\let\olddagger\dagger
\renewcommand{\dagger}{\ensuremath{\olddagger}\xspace}

\usepackage{makeidx}
\makeindex

\theoremstyle{plain}
\newtheorem*{main theorem}{Main Theorem}
\newtheorem{theorem}{Theorem}[section]
\newtheorem{corollary}[theorem]{Corollary}
\newtheorem{lemma}[theorem]{Lemma}

\newtheorem{definition}[theorem]{Definition}

\newtheorem{example*}[theorem]{Example*}
\newtheorem{examples*}[theorem]{Examples*}

\newtheorem{remark*}[theorem]{Remark*}

\newtheorem*{search problem}{Search Problem}

\usepackage{color}
\def\bR{\begin{color}{red}}
\def\bB{\begin{color}{blue}}
\def\bM{\begin{color}{magenta}}
\def\bC{\begin{color}{cyan}}
\def\bW{\begin{color}{white}}
\def\bBl{\begin{color}{black}}
\def\bG{\begin{color}{green}}
\def\bY{\begin{color}{yellow}}
\def\e{\end{color}\xspace}
\newcommand{\bit}{\begin{itemize}}
\newcommand{\eit}{\end{itemize}\par\noindent}
\newcommand{\ben}{\begin{enumerate}}
\newcommand{\een}{\end{enumerate}\par\noindent}
\newcommand{\beq}{\begin{equation}}
\newcommand{\eeq}{\end{equation}\par\noindent}
\newcommand{\beqa}{\begin{eqnarray*}}
\newcommand{\eeqa}{\end{eqnarray*}\par\noindent}
\newcommand{\beqn}{\begin{eqnarray}}
\newcommand{\eeqn}{\end{eqnarray}\par\noindent}

\def\jR{\begin{color}{black}}
\def\jB{\begin{color}{black}}
\def\jM{\begin{color}{magenta}}
\def\jC{\begin{color}{cyan}}
\def\jW{\begin{color}{white}}
\def\jBl{\begin{color}{black}}
\def\jG{\begin{color}{green}}
\def\jY{\begin{color}{yellow}}

\begin{document}

\begingroup
\centering
{\Large\textbf{A no-go theorem for theories that decohere to quantum mechanics} \\[1.5em]
 \normalsize  Ciar{\'a}n~M. Lee\textsuperscript{$\dagger$}\footnote{ciaran.lee@ucl.ac.uk} and John H. Selby\textsuperscript{$\ast$,}\footnote{jselby@perimeterinstitute.ca}
}
\\[1em]
\it \textsuperscript{$\dagger$} Department of Physics and Astronomy, University College London, Gower Street, London WC1E 6BT, UK. \\
 \it \textsuperscript{$\ast$} Department of Computer Science, University of Oxford, OX1 3QD, UK,\\ Department of Physics, Imperial College London,  London SW7 2AZ, UK,\\ Perimeter Institute for Theoretical Physics, Waterloo, Ontario, N2L 2Y5, Canada. \\

\endgroup

\begin{abstract}
{\color{black} To date, there has been no experimental evidence that invalidates quantum theory.} Yet it may only be an effective description of the world, in the same way that classical physics is an effective description of the quantum world. We ask whether there exists an operationally-defined theory superseding quantum theory, but which reduces to it via a decoherence-like mechanism. We prove that no such post-quantum theory exists if it is demanded that it satisfy two natural physical principles: \emph{causality} and \emph{purification}. Causality formalises the statement that information propagates from present to future, and purification that each state of incomplete information arises in an essentially unique way due to lack of information about an environment. Hence, our result can either be viewed as evidence that the fundamental theory of Nature is quantum, or as showing in a rigorous manner that any post-quantum theory must abandon causality, purification, or both.
\end{abstract}

\section{Introduction}
In 1903 Michelson wrote \emph{``The more important fundamental laws and facts of physical science have all been discovered, and these are so firmly established that the possibility of their ever being supplanted in consequence of new discoveries is exceedingly remote''} \cite{michelson1903light}. Within two years Einstein had proposed the photoelectric effect \cite{einstein1905photoelectric} and within thirty quantum theory was an established field of scientific research. This new science revolutionised our understanding of the physical world and brought with it a
{\color{black} host} of classically counter-intuitive features such as superposition, entanglement, and fundamental uncertainty.

{\color{black} Today, quantum theory has been verified to extremely high precision and
forms the basis of a vast array of new technologies.}
Yet, just as for Michelson, it may turn out to be the case that quantum theory is only an effective description of our world. There may be some more fundamental theory yet to be discovered that is as radical a departure from quantum theory as quantum was from classical. If such a theory exists, there should be some mechanism by which effects of this theory are suppressed, explaining why quantum theory is a good effective description of Nature. This would be analogous to decoherence, which both suppresses quantum effects and gives rise to the classical world \cite{joos2013decoherence, zurek2003decoherence,richens2017entanglement}. As such, this mechanism is called \emph{hyperdecoherence}. To the best of the authors knowledge, the notion of hyperdecoherence was first discussed in \cite{zyczkowski2008quartic} and has commonly been considered as a mechanism to explain why we do not observed post-quantum effects, such as in \cite{dakic2009quantum}, and, in particular, in the context of higher-order interference \cite{sorkin1994quantum,lee2016higher,lee2016deriving,lee2016generalised,barnum2014higher,bolotin2016ongoing,niestegge2013three,henson2015bounding,ududec2011three,sinha2015superposition,barnum2017ruling,lee2017oracles}.

We formalise such a hyperdecoherence mechanism within a broad framework of operationally-defined physical theories by generalising the key features of quantum to classical decoherence. Using this we prove a no-go result: there is no operationally-defined theory that satisfies two natural physical principles, \emph{causality} and \emph{purification}, and which reduces to quantum theory via a hyperdecoherence mechanism. Here, causality formalises the statement that information propagates from present to future, and purification that each state of incomplete information arises in an essentially unique way due to a lack of information about some larger environment system. In a sense, purification can be thought of as a statement of ``information conservation''; any missing information about the state of a given system can always be accounted for by considering it as part of a larger system. Our result can either be viewed as a justification of why the fundamental theory of Nature is quantum, or as highlighting the ways in which any post-quantum theory must radically depart from a quantum description of the world.

\section{Decoherence} \label{Decoherence example: classical from quantum}

One of the standard descriptions of the quantum to classical transition is environment-induced decoherence \cite{zurek2003decoherence}\footnote{For alternative approaches, see \cite{RKastner1, RKastner2} and references therein.}. In this description, a quantum system interacts deterministically with some environment system, after which the environment is discarded, leading to a loss of information. This procedure formalises the idea of a quantum system irretrievably losing information to an environment,  leading to an effective classical description of the decohered system. The decoherence process can be viewed as inducing a completely positive trace preserving map on the original quantum system, which is termed the \emph{decoherence map}.

A concrete example serves to illustrate the key features of this map. Consider the following reversible interaction with an environment:
$U=\sum_i \ketbra{i}{i} \otimes \pi_i, $
where $\{\ket{i}\}$ is the computational basis and $\pi_i$ is a unitary which acts on the environment system as $\pi_i\ket{0}=\ket{i} \text{, }\forall \text{ } i$. Switching to the density matrix formalism, the decoherence map arising from the above interaction corresponds to
$$\mathcal{D}(\rho)=\mathrm{Tr}_E\left(U\rho\otimes \ketbra{0}{0}_E U^\dagger \right)=\sum_i \bra{i} \rho \ket{i} \ketbra{i}{i},$$
where $\rho$ is the input state. Hence, in this concrete setting, the decoherence map $\mathcal{D}$ is a de-phasing map.

It is clear that $\mathcal{D}(\rho)$ will always be diagonal in the $\{\ket{i}\}$ basis, regardless of the input. Hence, as they have no coherences between distinct elements of $\{\ket{i}\}$, the states $\mathcal{D}(\rho)$ correspond to classical probability distributions. In fact, the entirety of classical probability theory---corresponding to probability distributions over classical outcomes, stochastic maps acting on said distributions, and measurements allowing one to infer the probabilities of different possible outcomes---can be seen to arise from quantum theory by applying $\mathcal{D}$ to density matrices $\rho$ as $\mathcal{D}(\rho)$, completely positive trace preserving maps $\mathcal{E}$ as $\mathcal{D}\left( \mathcal{E}\left(\mathcal{D}(\_)\right)\right)$, and POVM elements $M$ as $\mathrm{Tr}\left(M\mathcal{D}(\_)\right)$. In this manner, one can consider classical probability theory to be a sub-theory of quantum theory---meaning that applying stochastic maps to probability distributions results in probability distributions---where $\mathcal{D}$ is the map restricting quantum theory to the classical sub-theory. \color{black} The statement of the previous line is encompassed by what it meant by ``sub-theory''; as a sub-theory is itself a theory, it must be closed under composition.\color{black}

There are three key features of the decoherence map that we will use to define our hyper-decoherence map in section~\ref{Defining hyperdecoherence via thee key properties of decoherence}:
\begin{enumerate}
\item[i.] It is trace preserving, corresponding to the fact that it is a deterministic process.
\item[ii.] It is idempotent, meaning $$\mathcal{D}\left(\mathcal{D}(\rho)\right)=\mathcal{D}(\rho), \text{ for all } \rho.$$
    {\color{black}This corresponds to the intuitive fact that classical systems have no more coherence `to lose' and, moreover, once states have lost their coherence they are left invariant by further decoherence.}
\item[iii.] {\color{black}
    Finally, we observe that decoherence aises from an irretrievable loss of information to an environment, and so:
     \begin{enumerate}
     \item[a] If $\mathcal{D}(\rho)$ is a pure classical state, i.e. $\mathcal{D}(\rho)=\ketbra{i}{i}$ for some $i$, then $\rho$ is clearly also a pure quantum state. I.e. if the state that results from this loss of information is a state of maximal information, then no information can have been lost to the environment.
     \item[b] If $\mathcal{D}(\rho)$ is the maximally mixed classical state, i.e. $\mathcal{D}(\rho)=\frac{1}{d}\sum_{i=1}^d \ketbra{i}{i}$, then $\mathcal{D}(\rho)$ is clearly also a maximally mixed quantum state. I.e. if the decohered state is maximally ignorant regarding the classical state then it should be maximally ignorant about the quantum state.
     \end{enumerate}
     }
\end{enumerate}

\section{Generalised theories} \label{Generalised theories}

To make progress on the question raised at the start of this paper, we need to be able to describe theories other than quantum and classical in a consistent manner. This calls for a broad framework that can describe any conceivable---yet well-defined---physical theory. The framework we present here is based on \cite{coecke2016picturing,chiribella2010probabilistic,hardy2011reformulating,lee2015computation}\footnote{To be precise, one can view a theory in this framework as a generalised probabilistic theory \cite{chiribella2010probabilistic,hardy2011reformulating} where the standard assumptions of finite-dimensionality and closure of the set of states are not made, or as a process theory \cite{coecke2016picturing} with the added assumptions of tomography and convexity.} and takes the view that, ultimately, any physical theory is going to be explored by experiments, and so must have an \emph{operational} description in terms of these experiments.

Note that operationalism as a philosophical viewpoint, in which one asserts that there is no reality beyond laboratory device settings and outcomes, is not being espoused here. One should merely view the approach taken here as an operational methodology aimed at gaining insight into certain structural properties of physical theories. This operational approach {\color{black} is similar in spirit to that taken in device-independent quantum information processing---a field that has led to many practical applications \cite{barrett2005no, vazirani2014fully}\footnote{{\color{black}See \cite{chiribella2016bridging} for further details on the connection between these two frameworks.}}.}

A theory in this framework can be described as a collection of \emph{processes}, each of which corresponds to a particular outcome occurring in a single use of a piece of lab equipment in some experiment. In general, each process has some number of inputs and outputs. \color{black} These inputs and outputs are collectively called \emph{systems}. These systems are labelled by different \emph{types}, denoted $A,B,\dots$. Given two systems of type $A$ and $B$, we can form a \emph{composite system} of type $AB$. Operationally, a process with input system of type $AB$ corresponds to a single use of a piece of lab equipment with an input system of type $A$ and a distinct input system of type $B$. In finite-dimensional quantum theory, systems correspond to complex Hilbert spaces, with the type given by the dimension of the Hilbert space. Hence a type $A$ in quantum theory is just a natural number, that is $A\in\mathbb{N}$. Consider a qubit, which in our language corresponds to a quantum system of type $2$. Physically, a qubit can be realised in many different ways, such as by a spin-$1/2$ system or an ion in a trap with two distinct energy levels. Although these physical set-ups might differ, they are operationally equivalent. Hence, while the intuitive picture of a system corresponding to a particle ``passing from input to output port of a laboratory device'' is appealing, one should take care that this intuitive idea does not lead to ambiguities.

Processes with no inputs are known as \emph{states}---corresponding to density matrices in quantum theory---those with no outputs \emph{effects}---corresponding to POVM elements in quantum theory, and those having both inputs and outputs \emph{transformations}---corresponding to completely positive trace non-increasing maps in quantum theory.

The key feature of a theory in this framework is in how these processes compose. There are two equivalent ways to define this, diagrammatically or algebraically. Diagrammatically, we denote processes as labeled boxes and systems as labeled wires. Processes can then be wired together to form \emph{diagrams} such as:
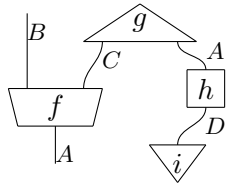
\[%
\begin{tikzpicture}
	\begin{pgfonlayer}{nodelayer}
		\node [style=none] (0) at (-1.75, 0.5000001) {};
		\node [style=none] (1) at (-1.5, -0.5000001) {};
		\node [style=none] (2) at (-0.5000001, -0) {$f$};
		\node [style=none] (3) at (0.7500001, 0.5000001) {};
		\node [style=none] (4) at (0.5000001, -0.5000001) {};
		\node [style=none] (5) at (-1.25, 0.5000001) {};
		\node [style=none] (6) at (-1.25, 2.5) {};
		\node [style=none] (7) at (0.25, 0.5000001) {};
		\node [style=none] (8) at (0.75, 1.75) {};
		\node [style=none] (9) at (-0.5000001, -0.5000001) {};
		\node [style=none] (10) at (-0.5000001, -1.5) {};
		\node [font={\footnotesize}, style=none] (11) at (-0.25, -1.25) {$A$};
		\node [font={\footnotesize}, style=none] (12) at (-1, 2) {$B$};
		\node [font={\footnotesize}, style=none] (13) at (1, 1.25) {$C$};
		\node [style=none] (14) at (0.25, 1.75) {};
		\node [style=none] (15) at (1.75, 2.75) {};
		\node [style=none] (16) at (3.25, 1.75) {};
		\node [style=none] (17) at (2.75, 1.75) {};
		\node [style=none] (18) at (0.75, 1.75) {};
		\node [style=none] (19) at (1.75, 2.25) {$g$};
		\node [style=none] (20) at (3, 1) {};
		\node [style=none] (21) at (3.5, 0.5) {$h$};
		\node [style=none] (22) at (4, 1) {};
		\node [style=none] (23) at (4, -0) {};
		\node [style=none] (24) at (3, -0) {};
		\node [style=none] (25) at (3.5, -0) {};
		\node [style=none] (26) at (2.75, -1) {};
		\node [style=none] (27) at (2, -1) {};
		\node [style=none] (28) at (2.75, -2) {};
		\node [style=none] (29) at (3.5, -1) {};
		\node [style=none] (30) at (2.75, -1.5) {$i$};
		\node [style=none] (31) at (3.5, 1) {};
		\node [font={\footnotesize}, style=none] (32) at (3.75, -0.5) {$D$};
		\node [font={\footnotesize}, style=none] (33) at (3.75, 1.5) {$A$};
	\end{pgfonlayer}
	\begin{pgfonlayer}{edgelayer}
		\draw (0.center) to (1.center);
		\draw (1.center) to (4.center);
		\draw (4.center) to (3.center);
		\draw (3.center) to (0.center);
		\draw (6.center) to (5.center);
		\draw [in=90, out=-90, looseness=1.00] (8.center) to (7.center);
		\draw (9.center) to (10.center);
		\draw (14.center) to (16.center);
		\draw (16.center) to (15.center);
		\draw (15.center) to (14.center);
		\draw (27.center) to (29.center);
		\draw (29.center) to (28.center);
		\draw (27.center) to (28.center);
		\draw [in=90, out=-90, looseness=1.00] (25.center) to (26.center);
		\draw (20.center) to (24.center);
		\draw (24.center) to (23.center);
		\draw (23.center) to (22.center);
		\draw (22.center) to (20.center);
		\draw [in=90, out=-90, looseness=1.00] (17.center) to (31.center);
	\end{pgfonlayer}
\end{tikzpicture}}\]
This wiring together of processes must satisfy two conditions: firstly, system types must match, and secondly, no cycles can be created. The relevant data for a particular diagram is just the \emph{connectivity}: which outputs are connected to which inputs and the ordering of the free inputs and outputs. Any circuits formed in this way must also correspond to a valid process in the theory. That is, for all theories in this framework, processes must be closed under this composition. Hence the above diagram must correspond to a process in the theory, in this case one with input system of type $A$ and output system of type $B$. One can think of the above diagram formed by connecting different processes as akin to circuits drawn in the field of quantum computation.

The equivalent algebraic statement formally corresponds to the fact that these systems and processes form the objects and morphisms of a strict symmetric monoidal category, see Ref.'s~\cite{coecke2016picturing} and \cite{chiribella2010probabilistic} for more information on the formal mathematical underpinnings of this. However, more intuitively, we can think of building the above diagrams out of two fundamental forms of composition, sequential and parallel. If $e$ is a process from a system of type $A$ to a system of type $B$ and $u$ is a process from system of type $B$ to system of type $C$, then their sequential composition is a process from a system of type $A$ to a system of type $C$, denoted $u\circ e$. Note that to sequentially compose two process, the type of the output system of the first process must match the type of the input system of the second. Similarly, if $e$ is a process from system of type $A$ to system of type $B$ and $u$ is a process from system of type $C$ to system of type $D$, then their parallel composition is a process from the composite system of type $AC$ to the composite system of type $BD$, denoted $u\otimes e$. Note that the symbol $\otimes$---which schematically denotes parallel composition---may not correspond to the standard vector space tensor product.

The definition of a strict symmetric monoidal category is then merely a statement that these two forms of composition interact in the way that one would expect, for example \cite{chiribella2010probabilistic, chiribella2011informational, coecke2016picturing}:
$$\big(u\otimes e\big)\circ\big( f\otimes k\big)=\big(u\circ f\big)\otimes\big(e\circ k\big)$$
for every process $u, e, f, k$ with the property that the type of the output system of $f$ (respectively, $k$) matches the type of the input system of $u$ (respectively, $e$). Note that this is exactly what happens in quantum theory.

To illustrate the connection between the algebraic and diagrammatic representation consider the above diagram translated into algebraic notation.
\begin{equation} %
\begin{tikzpicture}
	\begin{pgfonlayer}{nodelayer}
		\node [style=none] (0) at (-1.75, 0.5000001) {};
		\node [style=none] (1) at (-1.5, -0.5000001) {};
		\node [style=none] (2) at (-0.5000001, -0) {$f$};
		\node [style=none] (3) at (0.7500001, 0.5000001) {};
		\node [style=none] (4) at (0.5000001, -0.5000001) {};
		\node [style=none] (5) at (-1.25, 0.5000001) {};
		\node [style=none] (6) at (-1.25, 2.5) {};
		\node [style=none] (7) at (0.25, 0.5000001) {};
		\node [style=none] (8) at (0.75, 1.75) {};
		\node [style=none] (9) at (-0.5000001, -0.5000001) {};
		\node [style=none] (10) at (-0.5000001, -1.5) {};
		\node [font={\footnotesize}, style=none] (11) at (-0.25, -1.25) {$A$};
		\node [font={\footnotesize}, style=none] (12) at (-1, 2) {$B$};
		\node [font={\footnotesize}, style=none] (13) at (1, 1.25) {$C$};
		\node [style=none] (14) at (0.25, 1.75) {};
		\node [style=none] (15) at (1.75, 2.75) {};
		\node [style=none] (16) at (3.25, 1.75) {};
		\node [style=none] (17) at (2.75, 1.75) {};
		\node [style=none] (18) at (0.75, 1.75) {};
		\node [style=none] (19) at (1.75, 2.25) {$g$};
		\node [style=none] (20) at (3, 1) {};
		\node [style=none] (21) at (3.5, 0.5) {$h$};
		\node [style=none] (22) at (4, 1) {};
		\node [style=none] (23) at (4, -0) {};
		\node [style=none] (24) at (3, -0) {};
		\node [style=none] (25) at (3.5, -0) {};
		\node [style=none] (26) at (2.75, -1) {};
		\node [style=none] (27) at (2, -1) {};
		\node [style=none] (28) at (2.75, -2) {};
		\node [style=none] (29) at (3.5, -1) {};
		\node [style=none] (30) at (2.75, -1.5) {$i$};
		\node [style=none] (31) at (3.5, 1) {};
		\node [font={\footnotesize}, style=none] (32) at (3.75, -0.5) {$D$};
		\node [font={\footnotesize}, style=none] (33) at (3.75, 1.5) {$A$};
	\end{pgfonlayer}
	\begin{pgfonlayer}{edgelayer}
		\draw (0.center) to (1.center);
		\draw (1.center) to (4.center);
		\draw (4.center) to (3.center);
		\draw (3.center) to (0.center);
		\draw (6.center) to (5.center);
		\draw [in=90, out=-90, looseness=1.00] (8.center) to (7.center);
		\draw (9.center) to (10.center);
		\draw (14.center) to (16.center);
		\draw (16.center) to (15.center);
		\draw (15.center) to (14.center);
		\draw (27.center) to (29.center);
		\draw (29.center) to (28.center);
		\draw (27.center) to (28.center);
		\draw [in=90, out=-90, looseness=1.00] (25.center) to (26.center);
		\draw (20.center) to (24.center);
		\draw (24.center) to (23.center);
		\draw (23.center) to (22.center);
		\draw (22.center) to (20.center);
		\draw [in=90, out=-90, looseness=1.00] (17.center) to (31.center);
	\end{pgfonlayer}
\end{tikzpicture}} \quad = \quad \begin{array}{l}\left[\mathds{1}_B\otimes g^{C{A_2}}\right]\circ  \\
 \qquad \left[f^{A_1}_{BC}\otimes h^D_{A_2}\right]\circ \\ \qquad\qquad\left[\mathds{1}_{A_1}\otimes i_{D} \right]\end{array}
\end{equation}
where, on the right, $\mathds{1}$ corresponds to the identity process and $\otimes$ and $\circ$ denote parallel and sequential composition of processes respectively. In what follows, the $\circ$ will generally be suppressed. Algebraically, a process $d$ from system of type $A$ to system of type $B$, is denoted $d^A_B$. If the output system is of the same type as the input system, then the indices will be suppressed to a subscript, rather than a subscript and superscript. If there is no input/output system, the corresponding superscript/subscript is left blank. Note that in the right-hand algebraic equation, a dummy index on the repeated type $A$ had to be introduced as a book keeping measure, despite the fact $A_1$ and $A_2$ are the exact same type. Note the diagrammatic notation was able to deal with this issue without the need for a dummy index.

The following concrete example illustrates potential issues that can arise if one forgets the dummy index is merely a book keeping measure. Consider the quantum Bell state $\frac{1}{d}\sum_{ij}\ketbra{ii}{jj}$ in $d^2$ dimensions. As this is a maximally entangled two-qudit state, the type of each system is the same, namely the natural number $d$. However, in order to prevent ambiguity when marginalising over one of the qudit systems, we introduce a dummy index on the type, as follows:
\begin{equation}\label{marginalised Bell state}
 \mathrm{Tr}_{d_1}\left(\frac{1}{d}\sum_{ij} \ket{i}_{d_1}\bra{j}\otimes\ket{i}_{d_2}\bra{j} \right)=\frac{\mathds{1}_{d_2}}{d},
\end{equation}
where in the above $\otimes$ is the standard vector space tensor product. Clearly, marginalising over the other system results in
\begin{equation}\label{marginalised Bell state 2}
 \mathrm{Tr}_{d_2}\left(\frac{1}{d}\sum_{ij} \ket{i}_{d_1}\bra{j}\otimes\ket{i}_{d_2}\bra{j} \right)=\frac{\mathds{1}_{d_1}}{d}.
\end{equation}
As $d_1=d_2=d$, one has
\begin{equation}\label{marginalised Bell state 3}
\begin{aligned}
& \mathrm{Tr}_{d_1}\left(\frac{1}{d}\sum_{ij} \ket{i}_{d_1}\bra{j}\otimes\ket{i}_{d_2}\bra{j} \right) \\
 &=\frac{\mathds{1}}{d}=\mathrm{Tr}_{d_2}\left(\frac{1}{d}\sum_{ij} \ket{i}_{d_1}\bra{j}\otimes\ket{i}_{d_2}\bra{j} \right).
\end{aligned}
\end{equation}
That is, each marginalised state is the same, despite the fact that these systems can be space-like separated. It is the mathematical assignment of a state to each system that is the same, not the physical set-up. We saw above that in order marginalise correctly using algebraic notation, a dummy index had to be introduced to specify the system on which to apply the partial trace. However, it was important to note that after this procedure was completed, it was crucial to drop the dummy index.

When the circuit representing the connections of processes in an experiment has no free inputs or outputs, we associate it to the probability that all of these processes occur when the experiment is run, for example:
\[\textbf{Pr}(f,g,h,i)  \ \ := \ \ \ %
\begin{tikzpicture}
	\begin{pgfonlayer}{nodelayer}
		\node [style=none] (0) at (0, -0) {};
		\node [style=none] (1) at (0.7500001, -1.25) {};
		\node [style=none] (2) at (0.7500001, -0.5000001) {$f$};
		\node [style=none] (3) at (1.5, -0) {};
		\node [style=none] (4) at (0.7500001, -1.25) {};
		\node [style=none] (5) at (0.7500001, -0) {};
		\node [style=none] (6) at (1.5, 1.25) {};
		\node [style=none, font={\footnotesize}] (7) at (1.5, 0.5) {$C$};
		\node [style=none] (8) at (1, 1.25) {};
		\node [style=none] (9) at (2.5, 2.25) {};
		\node [style=none] (10) at (4, 1.25) {};
		\node [style=none] (11) at (3.5, 1.25) {};
		\node [style=none] (12) at (1.5, 1.25) {};
		\node [style=none] (13) at (2.5, 1.75) {$g$};
		\node [style=none] (14) at (4, 0.5000001) {};
		\node [style=none] (15) at (4.5, -0) {$h$};
		\node [style=none] (16) at (5, 0.5000001) {};
		\node [style=none] (17) at (5, -0.5000001) {};
		\node [style=none] (18) at (4, -0.5000001) {};
		\node [style=none] (19) at (4.5, -0.5000001) {};
		\node [style=none] (20) at (3.75, -1.25) {};
		\node [style=none] (21) at (2.25, -1.25) {};
		\node [style=none] (22) at (1.75, -1.25) {};
		\node [style=none] (23) at (3, -2.25) {};
		\node [style=none] (24) at (4.25, -1.25) {};
		\node [style=none] (25) at (2.5, 1.25) {};
		\node [style=none] (26) at (3, -1.75) {$i$};
		\node [style=none] (27) at (4.5, 0.5000001) {};
		\node [style=none, font={\footnotesize}] (28) at (2.75, 0.25) {$D$};
		\node [style=none, font={\footnotesize}] (29) at (4.75, -1) {$D$};
		\node [style=none, font={\footnotesize}] (30) at (4.75, 1) {$A$};
	\end{pgfonlayer}
	\begin{pgfonlayer}{edgelayer}
		\draw (0.center) to (1.center);
		\draw (1.center) to (4.center);
		\draw (4.center) to (3.center);
		\draw (3.center) to (0.center);
		\draw [in=90, out=-90, looseness=1.00] (6.center) to (5.center);
		\draw (8.center) to (10.center);
		\draw (10.center) to (9.center);
		\draw (9.center) to (8.center);
		\draw (22.center) to (24.center);
		\draw (24.center) to (23.center);
		\draw (22.center) to (23.center);
		\draw [in=90, out=-90, looseness=1.00] (19.center) to (20.center);
		\draw [in=90, out=-90, looseness=1.00] (25.center) to (21.center);
		\draw (14.center) to (18.center);
		\draw (18.center) to (17.center);
		\draw (17.center) to (16.center);
		\draw (16.center) to (14.center);
		\draw [in=90, out=-90, looseness=1.00] (11.center) to (27.center);
	\end{pgfonlayer}
\end{tikzpicture}} \]

There are two primitive experimental notions one would expect to be faithfully represented in any operationally-defined theory. The first is \emph{tomography}: if two processes give the same probabilities in all experiments, then they are the same process. That is:
\begin{equation} \label{tomography}
f=g \quad \iff \quad \forall X,\ \ \textbf{Pr}(f,X)=\textbf{Pr}(g,X)
\end{equation}
where $X$ is any diagram which, when composed with $f$ or $g$, has no free inputs and outputs. \color{black} Both quantum and classical theory actually satisfy the stronger notion of \emph{local tomography} where rather than quantifying over all $X$ we need only consider $X$ which are local state preparations and local effects. Note that this assumption is \emph{not} made for theories considered here. \color{black} The second is \emph{convexity}: given a collection of processes with the same inputs and outputs, experimentally it is possible to implement a probabilistic mixture of these, by applying one conditioned on the outcome of a coin toss for example. Hence a process corresponding to an arbitrary convex combination of processes, defined by
\begin{equation}\label{convexity}
h=\sum_ip_if_i  \iff \forall X, \textbf{Pr}(h,X)=\sum_ip_i \textbf{Pr}(f_i,X)
\end{equation}
where $p_i$ defines a probability distribution (i.e. $p_i\in \mathds{R}^+$ and $\sum_ip_i=1$), should exist in the theory.
Convexity allows us to define \emph{purity} of states. A state is \emph{pure} if it is not a convex combination of other distinct states. A state is \emph{mixed} if it can be written as a convex combination of distinct states.

From the above requirements, it can be shown that the set of states, effects, and transformations generate real vector spaces, with the effects and transformations acting linearly on the vector space of states \cite{chiribella2010probabilistic}.

\begin{definition}[Operational theory]\label{def:theory} A generalised theory consists of a collection of systems closed under parallel composition and processes closed under parallel and sequential composition, such that closed circuits formed from composing processes correspond to probability distributions. Moreover these processes satisfy \emph{tomography} and \emph{convexity} as defined in Eq.~(\ref{tomography}) and Eq.~(\ref{convexity}) respectively.
\end{definition}

In what follows, we will require our post-quantum theory to satisfy two natural physical principles, \emph{causality} and \emph{purification}, which were first introduced in \cite{chiribella2010probabilistic}. A process is \emph{deterministic} if the piece of lab equipment it corresponds to only has one possible outcome.

\begin{definition}[Causality \cite{chiribella2010probabilistic}]\label{def:causality} For each system of type $A$, there exists a unique deterministic  effect, {\color{black} denoted algebraically as, $ \mathsf{u}_A[\_]$, and diagrammatically as:
\[%
\begin{tikzpicture}
	\begin{pgfonlayer}{nodelayer}
		\node [style=upground] (0) at (0, 1) {};
		\node [style=none] (1) at (0, 0.7500001) {};
		\node [style=none] (2) at (0, 0.25) {};
		\node [style=none, font={\footnotesize}] (3) at (0.75, 0.75) {$A$};
	\end{pgfonlayer}
	\begin{pgfonlayer}{edgelayer}
		\draw (1.center) to (2.center);
	\end{pgfonlayer}
\end{tikzpicture}
}\]
}
\end{definition}
This may seem like a somewhat odd definition for causality, however it can be shown to be equivalent to the statement that future measurement choices do not effect current experiments \cite{chiribella2010probabilistic}. It also implies the no superluminal signalling principle \cite{coecke2014terminality} and provides a unique definition of marginalisation for multi-system states. A process  $f:A\to B$ is said to be {\color{black} \emph{terminal}} if {\color{black}
$
\mathsf{u}_B\left[f\left[\_\right]\right]=u_A[\_]
$
i.e. diagrammatically:
\[
\begin{tikzpicture}
	\begin{pgfonlayer}{nodelayer}
		\node [style=upground] (0) at (-0.2500001, 1) {};
		\node [style=none] (1) at (-0.2500001, 0.75) {};
		\node [style=none] (2) at (-0.2500001, 0.75) {};
		\node [style={small box}] (3) at (-0.2500001, -0) {$f$};
		\node [style=none] (4) at (-0.25, -1) {};
		\node [style=none] (5) at (1.5, -0) {$=$};
		\node [style=none] (6) at (2.75, 0.25) {};
		\node [style=none] (7) at (2.75, -0.25) {};
		\node [style=none] (8) at (2.75, 0.25) {};
		\node [style=upground] (9) at (2.75, 0.5) {};
		\node [style=none, font={\footnotesize}] (10) at (0.5, 0.75) {$A$};
		\node [style={right label}] (11) at (-0.25, -1) {$B$};
		\node [style=none, font={\footnotesize}] (12) at (3.5, 0.25) {$B$};
		\node [style=none] (13) at (0.75, 0.75) {};
	\end{pgfonlayer}
	\begin{pgfonlayer}{edgelayer}
		\draw (2.center) to (3);
		\draw (3) to (4.center);
		\draw (6.center) to (7.center);
	\end{pgfonlayer}
\end{tikzpicture}}
\]
}
In quantum theory the unique deterministic effect is provided by the (partial) trace, that is in the quantum case $\mathsf{u}_A[\_]=\mathrm{Tr}_A[\_]$, and so terminal transformations are precisely those that are trace preserving. It can be shown for general theories \cite{chiribella2015entanglement} that both reversible and deterministic transformations are {\color{black} terminal.}

\begin{definition}[Purification \cite{chiribella2010probabilistic}]\label{def:purification} For every state on a given system of type $A$, there exists a pure bipartite state on some composite system of type $AB$, such that the original state arises as a marginalisation of this pure bipartite state, {\color{black}
$
\rho_A = \mathsf{u}_B[\psi_{AB}]
$, i.e. diagrammatically:
\[
\begin{tikzpicture}
	\begin{pgfonlayer}{nodelayer}
		\node [style=upground] (0) at (3.75, 0.75) {};
		\node [style=none] (1) at (3.75, 0.5) {};
		\node [style=none] (2) at (3.75, -0) {};
		\node [style=none] (3) at (2.25, -0) {};
		\node [style=none] (4) at (2.25, 0.75) {};
		\node [style=point] (5) at (-1, -0.25) {$\rho$};
		\node [style=none] (6) at (-1, 0.75) {};
		\node [style=none] (7) at (0.5000001, -0) {$=$};
		\node [style=none] (8) at (3, -0.5000001) {$\psi$};
		\node [style=none] (9) at (3, -1) {};
		\node [style=none] (10) at (1.5, -0) {};
		\node [style=none] (11) at (4.5, -0) {};
		\node [style={right label}] (12) at (-1, 0.75) {$A$};
		\node [style={right label}] (13) at (2.25, 0.75) {$A$};
		\node [style=none, font={\footnotesize}] (14) at (4.5, 0.5) {$ B$};
	\end{pgfonlayer}
	\begin{pgfonlayer}{edgelayer}
		\draw (1.center) to (2.center);
		\draw (6.center) to (5);
		\draw (10.center) to (9.center);
		\draw (9.center) to (11.center);
		\draw (11.center) to (10.center);
		\draw (4.center) to (3.center);
	\end{pgfonlayer}
\end{tikzpicture}}
\]
}
Here, $\psi$ is said to \emph{purify} $\rho$. Moreover, any two pure states $\psi$ and $\psi'$ on the same system which purify the same state are connected by a reversible transformation, {\color{black}
$
\psi_{AB}=(\mathds{1}_A\otimes R_B)[\psi'_{AB}]
$, i.e. diagrammatically:
\[
\begin{tikzpicture}
	\begin{pgfonlayer}{nodelayer}
		\node [style=proj] (0) at (3.75, 1.1) {$R$};
		\node [style=none] (1) at (3.75, -0) {};
		\node [style=none] (2) at (2.25, -0) {};
		\node [style=none] (3) at (2.25, 2) {};
		\node [style=none] (4) at (0.5000001, -0) {$=$};
		\node [style=none] (5) at (3, -0.5000001) {$\psi'$};
		\node [style=none] (6) at (3, -1) {};
		\node [style=none] (7) at (1.5, -0) {};
		\node [style=none] (8) at (4.5, -0) {};
		\node [style=none] (9) at (-0.5000001, -0) {};
		\node [style=none] (10) at (-2.75, 1.25) {};
		\node [style=none] (11) at (-1.25, -0) {};
		\node [style=none] (12) at (-2, -1) {};
		\node [style=none] (13) at (-3.5, -0) {};
		\node [style=none] (14) at (-2, -0.5000001) {$\psi$};
		\node [style=none] (15) at (-1.25, 1.25) {};
		\node [style=none] (16) at (-2.75, -0) {};
		\node [style=none] (17) at (3.75, 2) {};
		\node [style={right label}] (18) at (-2.75, 1.25) {$A$};
		\node [style={right label}] (19) at (-1.25, 1.25) {$B$};
		\node [style={right label}] (20) at (2.25, 2) {$A$};
		\node [style={right label}] (21) at (3.75, 2) {$B$};
		\node [style={right label}] (22) at (3.75, 0.25) {$B$};
	\end{pgfonlayer}
	\begin{pgfonlayer}{edgelayer}
		\draw (0) to (1.center);
		\draw (7.center) to (6.center);
		\draw (6.center) to (8.center);
		\draw (8.center) to (7.center);
		\draw (3.center) to (2.center);
		\draw (15.center) to (11.center);
		\draw (13.center) to (12.center);
		\draw (12.center) to (9.center);
		\draw (9.center) to (13.center);
		\draw (10.center) to (16.center);
		\draw (17.center) to (0);
	\end{pgfonlayer}
\end{tikzpicture}}
\]
}
\end{definition}
If one considers a pure state to be a state of maximal information, the purification principle formalises the statement that each state of incomplete information arises in an essentially unique way due to a lack of information about an environment. In a sense, purification can be thought of as a statement of ``information conservation''; any missing information about the state of a system can always be traced back to lack of information of some environment system. Or, more succinctly: information can only be discarded, not destroyed \cite{chiribella2015conservation}.

The purification principle, in conjunction with another natural principles, implies many quantum information processing \cite{chiribella2010probabilistic} and computational primitives \cite{lee2016generalised}. Examples include teleportation, no information without disturbance, no-bit commitment \cite{chiribella2010probabilistic,sikora2018simple}, and the existence of reversible controlled transformations \cite{lee2016generalised}. Moreover, purification also leads to a well-defined notion of thermodynamics \cite{chiribella2015entanglement,chiribella2016entanglement,chiribella2016purity}.

Some concrete examples of theories in this framework serve to illustrate the terminology introduced in this section. As mentioned at different points above, finite-dimensional quantum theory belongs to our framework. Systems are given by complex Hilbert spaces, with the type of each system corresponding to the dimension of the corresponding Hilbert space, which in our case will always be a natural number. States are density matrices---that is, positive semi-definite Hermitian operators of unit trace acting on the underlying Hilbert space---transformations are completely positive trace preserving maps and effects are elements of positive operator valued measurements, or POVMs. The real vector space generated by the set of density matrices is given by the real vector space of Hermitian operators, spanned by the density matrices. Parallel composition of states in quantum theory takes a particularly neat form: a joint state of a composite system is a positive operator acting on the standard vector space tensor product of the Hilbert spaces associated with the individual systems. In particular, bi-partite quantum states can always be written as a real linear combination of product states.

Quantum theory satisfies both causality and purification. Indeed, to illustrate purification, it is enough to note that every mixed state on a finite-dimensional system $\sum_i p_i \ketbra{i}{i}$ can be purified to a state $\ketbra{\psi}{\psi}$, where $\ket{\psi}=\sum_i \sqrt{p_i}\ket{i}\ket{i}$, by the introduction of a suitable extra system. Moreover, any other purification $\ket{\phi}$ must satisfy $\ket{\psi}=\left(\mathbb{I}\otimes U\right)\ket{\phi}$  with $U$ a unitary transformation. Purification is standardly referred to by mathematicians as the Gelfand-Naimark-Segal construction \cite{gelfand1943}.

The classical theory of finite-dimensional probability distributions and stochastic processes is also an example of a specific theory in this framework. A system is associated with a real vector space with the type corresponding to the dimension of said vector space, which can be thought of as the number of discrete outcomes of some test on that system. In this work when ``classical theory'' is mentioned, this is what we mean.

Other interesting examples of generalised theories are Spekkens toy model \cite{spekkens2007evidence}; theories in which the set of states of a single system correspond to Euclidean hyperballs of dimension $n$ \cite{massar2015hyperdense,masanes2013existence} (the $n=3$ case of such theories corresponds to the Bloch ball of quantum theory); the theory colloquially known as `Boxworld' \cite{barrett2007information} containing states that exhibit Popescu-Rohrlich correlations which maximally violate the CHSH inequality, while respecting the no super-luminal signalling principle \cite{popescu1998causality}; and a class of theories which each have the same pure states and reversible transformations as quantum theory, but with different Born rule, mixed states, and measurements \cite{galley2016classification}. \color{black}The existence of such alternate theories allows for an investigation of the structural and information-theoretic properties of theories where different physical principles may hold. Indeed, the information processing and computational power of these alternative theories can be studied in a systematic way \cite{landscape, lee2015proofs, lee2016information, selby2018make, sikora2018simple}. The ambition of such investigations is to provide a deep understanding of the connections between physical principles and information-theoretic advantages in a theory-independent manner, and to perhaps shed light on the infamous quantum computational ``speed-up'' \cite{lee2016bounds}. \color{black}

One might wonder whether quantum field theory provides an example of a theory in this framework. Indeed, this remains a subject of ongoing investigation. See, in particular, Ref.'s \cite{oeckl2014first} and \cite{d2014derivation, bisio2015free}. This issue is mathematical rather than conceptual, indeed many authors take an operational point of view when working with quantum field theory, especially in the emerging field of relativistic quantum information \cite{fuentes2005alice, alsing2006entanglement}.
\color{black}

\section{Hyperdecoherence}\label{sec:Hyperdecoherence}\label{Defining hyperdecoherence via thee key properties of decoherence}

In section~\ref{Decoherence example: classical from quantum}, the quantum to classical transition was modelled by a decoherence map restricting quantum systems to classical ones. We can analogously model a post-quantum to quantum transition with a \emph{hyperdecoherence} map, represented {\color{black}algebraically as $\mathcal{D}$ and diagrammatically by $%
\begin{tikzpicture}
	\begin{pgfonlayer}{nodelayer}
		\node [style=proj] (0) at (0, -0) {};
		\node [style=none] (1) at (0, 0.5) {};
		\node [style=none] (2) at (0, -0.5) {};
	\end{pgfonlayer}
	\begin{pgfonlayer}{edgelayer}
		\draw (1.center) to (0);
		\draw (0) to (2.center);
	\end{pgfonlayer}
\end{tikzpicture}
}$}, which restricts post-quantum systems---described by a generalised theory, def.~\ref{def:theory} from section~\ref{Generalised theories}---to quantum ones \footnote{The formalisation of the idea of hyperdecoherence presented here is built on work presented in \cite{selby2016leaks,selby2017physical} where quantum to classical decoherence is discussed in terms of `leaks' in generalised process theories. A closely related definition of decoherence is given within the framework of categorical probabilistic theories in \cite{gogioso2017categorical}.}. We now adapt the three key features of decoherence outlined at the end of section~\ref{Decoherence example: classical from quantum} to this general setting, ending this section with a formal definition of a post-quantum theory.
{\color{black}
\begin{enumerate}
\item[i.] As in the quantum to classical transition, we think of this hyperdecoherence map as arising via some deterministic interaction with an environment system, after which the environment is discarded by marginalising with the unique deterministic effect. Hence, as with standard decoherence, hyperdecoherence can be thought of as an irretrievable loss of information to an environment. As deterministic processes are {\color{black} terminal}, the hyperdecoherence map should be \emph{terminal}:{\color{black}
\[
\begin{tikzpicture}
	\begin{pgfonlayer}{nodelayer}
		\node [style=upground] (0) at (0, 0.75) {};
		\node [style=none] (1) at (0, 0.5) {};
		\node [style=none] (2) at (0, 0.5) {};
		\node [style=proj] (3) at (0, -0) {};
		\node [style=none] (4) at (0, -0.5) {};
		\node [style=none] (5) at (1.5, -0) {$=$};
		\node [style=none] (6) at (2.75, 0.25) {};
		\node [style=none] (7) at (2.75, -0.25) {};
		\node [style=none] (8) at (2.75, 0.25) {};
		\node [style=upground] (9) at (2.75, 0.5) {};
		\node [style={right label}] (10) at (0.25, -0.25)
{$A$};
		\node [style=right label] (11) at (0.75, 0.5)
{$A$};
		\node [style=right label] (12) at (3.5, 0.25)
{$A$};
	\end{pgfonlayer}
	\begin{pgfonlayer}{edgelayer}
		\draw (2.center) to (3);
		\draw (3) to (4.center);
		\draw (6.center) to (7.center);
	\end{pgfonlayer}
\end{tikzpicture}}
\]
}
This is the analogue of point $i.$ from the end of section~\ref{Decoherence example: classical from quantum}.

\item[ii.] Moreover, hyperdecohering twice should be the same as hyperdecohering once, as the hyperdecohered system has no more `post-quantum-coherence' to `lose'. Hence this map should be \emph{idempotent}:{\color{black}
\[
\begin{tikzpicture}
	\begin{pgfonlayer}{nodelayer}
		\node [style=none] (0) at (0, 0.25) {};
		\node [style=none] (1) at (0, 0.25) {};
		\node [style=proj] (2) at (0, -0.375) {};
		\node [style=none] (3) at (0, -0.875) {};
		\node [style=none] (4) at (1.5, -0) {$=$};
		\node [style=none] (5) at (0, 0.75) {};
		\node [style=proj] (6) at (0, 0.375) {};
		\node [style=none] (7) at (0, -0) {};
		\node [style=none] (8) at (0, 0.875) {};
		\node [style=none] (9) at (2.5, 0.5) {};
		\node [style=proj] (10) at (2.5, -0) {};
		\node [style=none] (11) at (2.5, -0.5) {};
		\node [style=none] (12) at (2.5, 0.5) {};
		\node [style={right label}] (13) at (0.25, 0.25) {$A$};
		\node [style={right label}] (14) at (0.25, -0.5) {$A$};
		\node [style={right label}] (15) at (2.75, -0.25) {$A$};
	\end{pgfonlayer}
	\begin{pgfonlayer}{edgelayer}
		\draw (1.center) to (2);
		\draw (2) to (3.center);
		\draw (8.center) to (6);
		\draw (6) to (7.center);
		\draw (12.center) to (10);
		\draw (10) to (11.center);
	\end{pgfonlayer}
\end{tikzpicture}}
\]
}
This is the analogue of point $ii.$ from the end of section~\ref{Decoherence example: classical from quantum}, where idempotence immediately followed from the fact that the decoherence map sends off-diagonal terms in the density matrix to zero, losing all quantum coherences in the process. A natural extension of quantum theory that has been considered is the theory of Density Cubes \cite{dakic2014density}, where states are rank-$3$ tensors satisfying some positivity conditions, rather than rank-$2$ density matrices. In this case, one can identify the `post-quantum-coherences' as the elements with three distinct indices. Hyperdecoherence would then correspond to sending these terms to zero, resulting in standard density matrices \cite{lee2016higher}. Such a procedure would again clearly be idempotent.

\item[iiia.]
One can define a notion of purity relative to the sub-theory constructed via the above procedure. A state from the sub-theory is \emph{pure in the sub-theory} if it cannot be written as a convex combination of other states \emph{from the sub-theory}. Note that a state which is pure in the sub-theory may not be pure in the full post-quantum theory, as a state that cannot be written as a convex combination of states from the sub-theory may turn out to be decomposable as a convex combination of states lying outside the sub-theory. As hyperdecoherence arises from an irretrievable loss of information to an environment, if a state resulting from this process is a state of maximal information, then no information can have been lost to the environment. We formalise this by demanding that pure states in the sub-theory are pure in the post-quantum theory. This is the analogue of point $iiia.$ from the end of section~\ref{Decoherence example: classical from quantum}.
\item[iiib.]
Similarly we can define the notion of a maximally mixed state relative to the sub-theory. A state from the sub-theory is \emph{maximally mixed in the sub-theory} if any state \emph{from the sub-theory} appears in some convex decomposition. Note that a state which is maximally mixed in the sub-theory may not be maximally mixed in the full theory. However, this would describe a rather odd situation where, under hyperdecoherence, the maximally mixed state from the full theory is mapped to a state containing more information. This is clearly in conflict with the idea that hyperdecoherence represents a loss of information to the environment. Hence, we demand that the state that is maximally mixed in the sub-theory is maximally mixed in the full theory. This is the analogue of point $iiib.$ from the end of section~\ref{Decoherence example: classical from quantum}.
\end{enumerate}
}

As was the case for classical theory in section~\ref{Decoherence example: classical from quantum}, one can construct the entirety of quantum theory as a sub-theory of the post-quantum theory by appropriately applying $\mathcal{D}$ to states, transformations, and effects from the post-quantum theory. That is, density matrices, completely positive trace non-increasing maps, and POVM elements correspond to {\color{black}
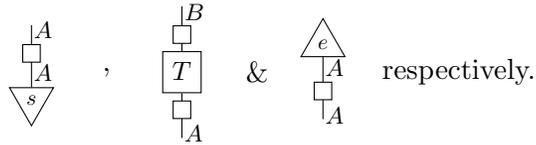
\[
\begin{tikzpicture}
	\begin{pgfonlayer}{nodelayer}
		\node [style=point] (0) at (0, -0.75) {$s$};
		\node [style=proj] (1) at (0, 0.5) {};
		\node [style=none] (2) at (0, 1.25) {};
		\node [style={small box}] (3) at (4, -0) {$T$};
		\node [style=proj] (4) at (4, 1) {};
		\node [style=proj] (5) at (4, -1) {};
		\node [style=none] (6) at (4, -1.75) {};
		\node [style=none] (7) at (4, 1.75) {};
		\node [style={right label}] (8) at (4, -1.75) {$A$};
		\node [style={right label}] (9) at (4, 1.5) {$B$};
		\node [style={right label}] (10) at (7.75, -1.25) {$A$};
		\node [style=copoint] (11) at (7.75, 0.75) {$e$};
		\node [style={right label}] (12) at (7.75, -0) {$A$};
		\node [style=proj] (13) at (7.75, -0.5) {};
		\node [style=none] (14) at (7.75, -1.25) {};
		\node [style=none] (15) at (2, -0) {,};
		\node [style=none] (16) at (6, -0) {\&};
		\node [style={right label}] (17) at (0, 1) {$A$};
		\node [style={right label}] (18) at (0, -0.125) {$A$};
		\node [style=none] (19) at (0, 1) {};
	\end{pgfonlayer}
	\begin{pgfonlayer}{edgelayer}
		\draw (2.center) to (1);
		\draw (1) to (0);
		\draw (7.center) to (4);
		\draw (4) to (3);
		\draw (3) to (5);
		\draw (5) to (6.center);
		\draw (14.center) to (13);
		\draw (13) to (11);
	\end{pgfonlayer}
\end{tikzpicture}}\quad \text{respectively.}
\]
}

Hence---as  $\mathcal{D}$ is idempotent---quantum states, transformations, and effects are those left invariant by the hyperdecoherence map. Note that, as in the quantum to classical case, a sub-theory is itself a theory and so it must be closed under both sequential and parallel composition.

{\color{black} Point iiia. above} will play an important role in our proof, so it is worth discussing in more detail here. Firstly note that we need some assumption in addition to {\color{black} terminality} and idempotence in order to capture a sensible notion of hyperdecoherence. Indeed, even to adequately capture the standard notion of decoherence, one needs constraints beyond {\color{black} terminality} and idempotence. To see this, consider the following example. Consider a system in classical probability theory of type $\mathcal{C}$. Define systems in a ``post-classical theory'' by tensoring two systems of type  $\mathcal{C}$ together to form a composite system of type {\color{black}$\mathbf{C}:=\mathcal{C} \otimes \mathcal{C}$}, with the decoherence map given by tracing out one of the systems and preparing a mixed classical state $\textbf{q}=\sum_i p_i \ket{i}\bra{i}$, such that $p_i >0$ for at least two distinct values of $i$, in its place. That is, here, {\color{black}$\mathcal{D}_\mathbf{C}:=\mathds{1}_{\mathcal{C}}\otimes (\textbf{q}\circ \mathrm{Tr}_\mathcal{C}(\_))$, or, diagrammatically:
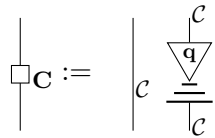
\[
\begin{tikzpicture}
	\begin{pgfonlayer}{nodelayer}
		\node [style=proj] (0) at (0, -0) {};
		\node [style={right label}] (1) at (0.25, -0.25) {$\mathbf{C}$};
		\node [style=none] (2) at (0, 1.5) {};
		\node [style=none] (3) at (0, -1.5) {};
		\node [style=none] (4) at (1.5, -0) {$:=$};
		\node [style=none] (5) at (3, 1.5) {};
		\node [style=none] (6) at (3, -1.5) {};
		\node [style={right label}] (7) at (3, -0.5) {$\mathcal{C}$};
		\node [style=upground] (8) at (4.5, -0.5) {};
		\node [style=none] (9) at (4.5, -0.75) {};
		\node [style=none] (10) at (4.5, -1.5) {};
		\node [style=point] (11) at (4.5, 0.5) {$\mathbf{q}$};
		\node [style=none] (12) at (4.5, 1.5) {};
		\node [style={right label}] (13) at (4.5, 1.5) {$\mathcal{C}$};
		\node [style={right label}] (14) at (4.5, -1.5) {$\mathcal{C}$};
	\end{pgfonlayer}
	\begin{pgfonlayer}{edgelayer}
		\draw (2.center) to (0);
		\draw (0) to (3.center);
		\draw (5.center) to (6.center);
		\draw (9.center) to (10.center);
		\draw (12.center) to (11);
	\end{pgfonlayer}
\end{tikzpicture}
\]}
 it is easy to see that this decoherence map is trace preserving (i.e. {\color{black} terminal}), idempotent, and recovers all states of the original $\mathcal{C}$ system---albeit tensored with a fixed mixed state. However, this does not properly capture the standard notion of decoherence as the ``post-classical theory'' is nothing but classical theory itself. Moreover, we can do a similar thing for quantum theory by having a quantum system of type $\mathcal{Q}$ that ``hyperdecoheres'' from the quantum composite system of type $\mathcal{Q}\otimes\mathcal{Q}$, such that the ``post-quantum theory'' is nothing but quantum theory itself.

Note that these examples are ruled out by our assumption that pure states in the decohered sub-theory are pure in the full theory. Indeed, applying $\mathcal{D}$ to the pure classical state $\textbf{a}\otimes \textbf{b}$, results in $$\textbf{a}\otimes\textbf{q}=\sum_i p_i \textbf{a} \otimes \ket{i}\bra{i},$$ but $\textbf{a}\otimes\ket{i}\bra{i}$ is not a state in the decohered sub-theory for any $i$. Hence in the sub-theory $\textbf{a}\otimes\textbf{q}$ is pure, but in the full theory it is not.

One might ask whether requiring that pure decohered states are pure in the full theory is the minimal assumption needed to rule out these examples. Indeed, demanding the seemingly weaker constraint that the pre- and post-decohered systems have the same dimension also rules them out. Phrased in operational terms, preserving the dimension corresponds to the hyperdecoherence map preserving the number of perfectly distinguishable states. This requirement rules out the above example. Indeed, if the decohered system has $n$ distinguishable states then the original system has $n^2$. However, we prove in appendix~\ref{Proof that pure quantum states are pure} that---given a strengthened version of purification---one can derive the requirement that pure quantum states are pure post-quantum states from the assumption that hyperdecoherences preserves the number of perfectly distinguishable states. This, in conjunction with the fact that pure classical states are always pure quantum states, leads us to propose the requirement that pure quantum states are pure as a defining feature of hyperdecoherence. \color{black} There is, however, a suggestion arising from insights into quantum gravity \cite{muller2009does}, that on a fundamental level pure quantum states may not actually be pure. We therefore see the need for this assumption as a feature rather than a bug as it lends further evidence to this assertion.
\color{black} See section~\ref{sec:Discussion} for a further rumination on this point.

A final requirement of hyperdecoherence is that the original theory is not the same theory as the decohered theory, that is, one of the hyperdecoherence maps must be non-trivial. We say a hyperdecoherence map is \emph{trivial} if it is equal to the identity transformation:{\color{black}
\[
%
\begin{tikzpicture}
	\begin{pgfonlayer}{nodelayer}
		\node [style=none] (0) at (0, -0) {};
		\node [style=none] (1) at (0, -1.5) {};
		\node [style=proj] (2) at (0, -0.75) {};
		\node [style=none] (3) at (1.5, -0.75) {$=$};
		\node [style=none] (4) at (2.5, -0) {};
		\node [style=none] (5) at (2.5, -1.5) {};
		\node [style={right label}] (6) at (0.25, -1) {$A$};
		\node [style={right label}] (7) at (2.5, -1) {$A$};
	\end{pgfonlayer}
	\begin{pgfonlayer}{edgelayer}
		\draw (0.center) to (2);
		\draw (2) to (1.center);
		\draw (4.center) to (5.center);
	\end{pgfonlayer}
\end{tikzpicture}}
\]
}

To summarise all of the above, we now formally define a post-quantum theory.

\begin{definition}[Post-quantum theory]\label{def:PQT} An operational theory (def.~\ref{def:theory}) is a post-quantum theory if, for each system of type $A$, there exists a hyperdecoherence map {\color{black}$%
\begin{tikzpicture}
	\begin{pgfonlayer}{nodelayer}
		\node [style=proj] (0) at (-0.25, -0) {};
		\node [style=none] (1) at (-0.25, 0.5) {};
		\node [style=none] (2) at (-0.25, -0.5) {};
		\node [style={right label}] (3) at (0, -0.25) {$A$};
	\end{pgfonlayer}
	\begin{pgfonlayer}{edgelayer}
		\draw (1.center) to (0);
		\draw (0) to (2.center);
	\end{pgfonlayer}
\end{tikzpicture}
}$\hspace{-2mm}}
satisfying the following
\begin{enumerate}
\item
{\color{black}$%
\begin{tikzpicture}
	\begin{pgfonlayer}{nodelayer}
		\node [style=proj] (0) at (-0.25, -0) {};
		\node [style=none] (1) at (-0.25, 0.5) {};
		\node [style=none] (2) at (-0.25, -0.5) {};
		\node [style={right label}] (3) at (0, -0.25) {$A$};
	\end{pgfonlayer}
	\begin{pgfonlayer}{edgelayer}
		\draw (1.center) to (0);
		\draw (0) to (2.center);
	\end{pgfonlayer}
\end{tikzpicture}
}$\hspace{-2mm}} is {\color{black} terminal}:  {\color{black}$%
\begin{tikzpicture}
	\begin{pgfonlayer}{nodelayer}
		\node [style=upground] (0) at (0, 0.75) {};
		\node [style=none] (1) at (0, 0.5) {};
		\node [style=none] (2) at (0, 0.5) {};
		\node [style=proj] (3) at (0, -0) {};
		\node [style=none] (4) at (0, -0.5) {};
		\node [style=none] (5) at (1.5, -0) {$=$};
		\node [style=none] (6) at (2.75, 0.25) {};
		\node [style=none] (7) at (2.75, -0.25) {};
		\node [style=none] (8) at (2.75, 0.25) {};
		\node [style=upground] (9) at (2.75, 0.5) {};
		\node [style={right label}] (10) at (0.25, -0.25)
{$A$};
		\node [style=right label] (11) at (0.75, 0.5)
{$A$};
		\node [style=right label] (12) at (3.5, 0.25)
{$A$};
	\end{pgfonlayer}
	\begin{pgfonlayer}{edgelayer}
		\draw (2.center) to (3);
		\draw (3) to (4.center);
		\draw (6.center) to (7.center);
	\end{pgfonlayer}
\end{tikzpicture}}$}
\item
{\color{black}$%
\begin{tikzpicture}
	\begin{pgfonlayer}{nodelayer}
		\node [style=proj] (0) at (-0.25, -0) {};
		\node [style=none] (1) at (-0.25, 0.5) {};
		\node [style=none] (2) at (-0.25, -0.5) {};
		\node [style={right label}] (3) at (0, -0.25) {$A$};
	\end{pgfonlayer}
	\begin{pgfonlayer}{edgelayer}
		\draw (1.center) to (0);
		\draw (0) to (2.center);
	\end{pgfonlayer}
\end{tikzpicture}
}$\hspace{-2mm}} is idempotent:
{\color{black}$%
\begin{tikzpicture}
	\begin{pgfonlayer}{nodelayer}
		\node [style=none] (0) at (0, 0.25) {};
		\node [style=none] (1) at (0, 0.25) {};
		\node [style=proj] (2) at (0, -0.375) {};
		\node [style=none] (3) at (0, -0.875) {};
		\node [style=none] (4) at (1.5, -0) {$=$};
		\node [style=none] (5) at (0, 0.75) {};
		\node [style=proj] (6) at (0, 0.375) {};
		\node [style=none] (7) at (0, -0) {};
		\node [style=none] (8) at (0, 0.875) {};
		\node [style=none] (9) at (2.5, 0.5) {};
		\node [style=proj] (10) at (2.5, -0) {};
		\node [style=none] (11) at (2.5, -0.5) {};
		\node [style=none] (12) at (2.5, 0.5) {};
		\node [style={right label}] (13) at (0.25, 0.25) {$A$};
		\node [style={right label}] (14) at (0.25, -0.5) {$A$};
		\node [style={right label}] (15) at (2.75, -0.25) {$A$};
	\end{pgfonlayer}
	\begin{pgfonlayer}{edgelayer}
		\draw (1.center) to (2);
		\draw (2) to (3.center);
		\draw (8.center) to (6);
		\draw (6) to (7.center);
		\draw (12.center) to (10);
		\draw (10) to (11.center);
	\end{pgfonlayer}
\end{tikzpicture}}$}
\item {\color{black}
\begin{enumerate}
\item[a.] Pure states in the sub-theory are pure states\footnote{See point iiia. at the start of this section}.
\item[b.] The maximally mixed state in the sub-theory is maximally mixed in the full theory\footnote{See point iiib. at the start of this section}.
\end{enumerate}}
\end{enumerate}
Moreover, the collection {\color{black}$\{
$%
\begin{tikzpicture}
	\begin{pgfonlayer}{nodelayer}
		\node [style=proj] (0) at (-0.25, -0) {};
		\node [style=none] (1) at (-0.25, 0.5) {};
		\node [style=none] (2) at (-0.25, -0.5) {};
		\node [style={right label}] (3) at (0, -0.25) {$A$};
	\end{pgfonlayer}
	\begin{pgfonlayer}{edgelayer}
		\draw (1.center) to (0);
		\draw (0) to (2.center);
	\end{pgfonlayer}
\end{tikzpicture}
}\hspace{-2mm}$
\}$} defines a sub-theory
which corresponds to quantum theory, and at least one of the hyperdecoherence maps must be non-trivial.
\end{definition}

\section{Main Result}\label{sec:Results}

\begin{main theorem}
There is no post-quantum theory (def.~\ref{def:PQT}) satisfying both causality (def.~\ref{def:causality}) and purification (def.~\ref{def:purification}).
\end{main theorem}
Before we present the proof, we give an intuitive sketch of how it will proceed. We prove that in any post-quantum theory satisfying causality and purification, the hyperdecoherence map must be trivial for all systems. The main idea of the proof is to show that by performing a suitable post-quantum measurement on the quantum Bell state and post-selecting on a suitable post-quantum effect, any post-quantum state can be steered to. As quantum states are left invariant by the hyperdecoherence map (even locally, as we show below), all post-quantum states are left invariant as well---due to the fact that they can be steered to using a quantum state. Hence, for each system, the hyperdecoherence map must be the identity, a contradiction.

{\color{black}
 We will now present a purely diagrammatic proof of the Main Theorem. However, for readers unfamiliar with diagrammatic notation, we will also provide a proof using standard algebraic notation in appendix~\ref{app:DiagrammaticProof}.}

\begin{proof}
For convenience we denote quantum states with a subscript $q$. {\color{black}As discussed at the end of section~\ref{Generalised theories}, given} a bipartite quantum state $\psi_q$, it can be written as
\[%
\begin{tikzpicture}
	\begin{pgfonlayer}{nodelayer}
		\node [style=none] (0) at (-0.75, -0) {};
		\node [style=none] (1) at (0, -0) {};
		\node [style=none] (2) at (1.5, -0) {};
		\node [style=none] (3) at (2.25, -0) {};
		\node [style=none] (4) at (0.7500001, -1.25) {};
		\node [style=none] (5) at (0.75, -0.5) {$\psi_q$};
		\node [style=none] (6) at (0, 0.5000001) {};
		\node [style=none] (7) at (1.5, 0.5000001) {};
		\node [style=none] (8) at (3, -0) {$=$};
		\node [style=point] (9) at (6.25, -0.5000001) {{\color{white}$\phi_q^i\ $}};
		\node [style=point] (10) at (8.000001, -0.5000001) {{\color{white}$\phi_q^i\ $}};
		\node [style=none] (11) at (4.5, -0.25) {$\sum\limits_{ij}r_{ij}$};
		\node [style=none] (12) at (6.25, 0.5000001) {};
		\node [style=none] (13) at (8.000001, 0.5000001) {};
		\node [style=none] (14) at (6.25, -0.5000001) {$\phi_q^i$};
		\node [style=none] (15) at (8.000001, -0.5000001) {$\chi_q^j$};
		\node [style=none] (16) at (11, -0) {$r_{ij}\in \mathds{R}$};
	\end{pgfonlayer}
	\begin{pgfonlayer}{edgelayer}
		\draw (0.center) to (3.center);
		\draw (3.center) to (4.center);
		\draw (4.center) to (0.center);
		\draw (6.center) to (1.center);
		\draw (7.center) to (2.center);
		\draw (12.center) to (9);
		\draw (13.center) to (10);
	\end{pgfonlayer}
\end{tikzpicture}}.\]
{\color{black}Where the fact that this holds even when representing quantum theory as a sub-theory of the post-quantum theory follows immediately from, i) the definition of a sub-theory, and ii) linearity of transformations.} Idempotence of the hyperdecoherence map (point $2.$ of def.~(\ref{def:PQT})) then gives
\beq \label{localmap1} %
\begin{tikzpicture}
	\begin{pgfonlayer}{nodelayer}
		\node [style=none] (0) at (0.25, -0) {};
		\node [style=none] (1) at (0.75, -0) {};
		\node [style=none] (2) at (2.25, -0) {};
		\node [style=none] (3) at (2.75, -0) {};
		\node [style=none] (4) at (1.5, -1.25) {};
		\node [style=none] (5) at (1.5, -0.5) {$\psi_q$};
		\node [style=none] (6) at (0.75, 1) {};
		\node [style=none] (7) at (3.5, -0) {$=$};
		\node [style=point] (8) at (7, -0.5) {{\color{white}$\phi_q^i\ $}};
		\node [style=point] (9) at (8.75, -0.5) {{\color{white}$\phi_q^i\ $}};
		\node [style=none] (10) at (5, -0.25) {$\sum\limits_{ij}r_{ij}$};
		\node [style=none] (11) at (7, 1) {};
		\node [style=none] (12) at (7, -0.5) {$\phi_q^i$};
		\node [style=none] (13) at (8.75, -0.5) {$\chi_q^j$};
		\node [style=proj] (14) at (2.25, 0.5) {};
		\node [style=none] (15) at (2.25, 1) {};
		\node [style=none] (16) at (8.75, -0.25) {};
		\node [style=none] (17) at (8.75, 1) {};
		\node [style=proj] (18) at (8.75, 0.5) {};
		\node [style=none] (19) at (11.5, 0.75) {};
		\node [style=none] (20) at (12.25, -1.25) {};
		\node [style=none] (21) at (11, -0) {};
		\node [style=none] (22) at (11.5, -0) {};
		\node [style=none] (23) at (13.5, -0) {};
		\node [style=none] (24) at (13, -0) {};
		\node [style=none] (25) at (12.25, -0.5) {$\psi_q$};
		\node [style=none] (26) at (10.25, -0) {$=$};
		\node [style=none] (27) at (13, 0.75) {};
	\end{pgfonlayer}
	\begin{pgfonlayer}{edgelayer}
		\draw (0.center) to (3.center);
		\draw (3.center) to (4.center);
		\draw (4.center) to (0.center);
		\draw (6.center) to (1.center);
		\draw (11.center) to (8);
		\draw (15.center) to (14);
		\draw (14) to (2.center);
		\draw (17.center) to (18);
		\draw (18) to (16.center);
		\draw (21.center) to (23.center);
		\draw (23.center) to (20.center);
		\draw (20.center) to (21.center);
		\draw (19.center) to (22.center);
		\draw (27.center) to (24.center);
	\end{pgfonlayer}
\end{tikzpicture}}\eeq

Next, consider the maximally mixed quantum state, $\mu_q:=\frac{\mathds{1}}{d}$, of a $d$-dimensional system, {\color{black} and note that, from point $3b.$ of def.~(\ref{def:PQT})), this is also maximally mixed for the post-quantum theory, hence for any  pure state $\phi$ we can write:

\beq
\label{eq:anyDecomposition1}
\begin{tikzpicture}
	\begin{pgfonlayer}{nodelayer}
		\node [style=point] (0) at (-6.5, -0.25) {$\mu_q$};
		\node [style=none] (1) at (-5, -0) {$=$};
		\node [style=none] (2) at (-6.5, 0.5) {};
		\node [style=point] (3) at (-2.5, -0.25) {$\phi$};
		\node [style=none] (4) at (0.25, -0) {$(1-\frac{1}{d})$};
		\node [style=none] (5) at (-2.5, 0.75) {};
		\node [style=none] (6) at (-3.5, -0) {$\frac{1}{d}$};
		\node [style=none] (7) at (2, 0.75) {};
		\node [style=none] (8) at (-1.25, -0) {$+$};
		\node [style=point] (9) at (2, -0.25) {$\sigma$};
	\end{pgfonlayer}
	\begin{pgfonlayer}{edgelayer}
		\draw (2.center) to (0);
		\draw (5.center) to (3);
		\draw (7.center) to (9);
	\end{pgfonlayer}
\end{tikzpicture}
\eeq
that is} \emph{any} pure state from the post-quantum theory arises in a decomposition of the quantum maximally mixed state.

{\color{black}Recall that} every (non-trivial) quantum system of type $A$ has at least two perfectly distinguishable states, $\{0_q:=\ketbra{0}{0},1_q:=\ketbra{1}{1}\}$. Given the decomposition of Eq.~(\ref{eq:anyDecomposition1}), \emph{convexity} (Eq.~(\ref{convexity})) implies the following is a state in the post-quantum theory:
\[%
\begin{tikzpicture}
	\begin{pgfonlayer}{nodelayer}
		\node [style=none] (0) at (4.75, -0) {$:=$};
		\node [style=point] (1) at (7, -0.25) {$\phi$};
		\node [style=none] (2) at (10.75, -0) {$(1-\frac{1}{d})$};
		\node [style=none] (3) at (7, 0.7499999) {};
		\node [style=none] (4) at (6, -0) {$\frac{1}{d}$};
		\node [style=none] (5) at (12.5, 0.75) {};
		\node [style=none] (6) at (9.25, -0) {$+$};
		\node [style=point] (7) at (12.5, -0.25) {$\sigma$};
		\node [style=none] (8) at (1, -0) {};
		\node [style=none] (9) at (4, -0) {};
		\node [style=none] (10) at (1.75, -0) {};
		\node [style=none] (11) at (3.25, -0) {};
		\node [style=none] (12) at (1.75, 0.7499999) {};
		\node [style=none] (13) at (3.25, 0.7499999) {};
		\node [style=none] (14) at (2.5, -0.5000001) {$s_\phi$};
		\node [style=none] (15) at (2.5, -0.9999999) {};
		\node [style={right label}] (16) at (1.75, 0.75) {$A$};
		\node [style={right label}] (17) at (3.25, 0.75) {$A$};
		\node [style=point] (18) at (8.25, -0.25) {$0_q$};
		\node [style=point] (19) at (13.75, -0.25) {$1_q$};
		\node [style=none] (20) at (8.25, 0.7499999) {};
		\node [style=none] (21) at (13.75, 0.75) {};
	\end{pgfonlayer}
	\begin{pgfonlayer}{edgelayer}
		\draw (12.center) to (10.center);
		\draw (13.center) to (11.center);
		\draw (8.center) to (15.center);
		\draw (15.center) to (9.center);
		\draw (9.center) to (8.center);
		\draw (20.center) to (18);
		\draw (21.center) to (19);
		\draw (3.center) to (1);
		\draw (5.center) to (7);
	\end{pgfonlayer}
\end{tikzpicture}}\]
Consider a purification of this state, denoted $\mathcal{S}_\phi$, and note that it has the following properties:
\ben
\item $%
\begin{tikzpicture}
	\begin{pgfonlayer}{nodelayer}
		\node [style=none] (0) at (4, -0) {$=$};
		\node [style=none] (1) at (4.75, -0) {};
		\node [style=none] (2) at (7.75, -0) {};
		\node [style=none] (3) at (5.5, -0) {};
		\node [style=none] (4) at (7, -0) {};
		\node [style=none] (5) at (5.5, 1) {};
		\node [style=none] (6) at (7, 1) {};
		\node [style=none] (7) at (6.25, -0.5000001) {$s_\phi$};
		\node [style=none] (8) at (6.25, -0.9999999) {};
		\node [style={right label}] (9) at (5.5, 0.5) {$A$};
		\node [style={right label}] (10) at (7, 0.5) {$A$};
		\node [style=none] (11) at (1.25, -0.9999999) {};
		\node [style=none] (12) at (-0.7500001, -0) {};
		\node [style={right label}] (13) at (1.25, 0.5) {$A$};
		\node [style=none] (14) at (1.25, 0.9999999) {};
		\node [style=none] (15) at (3.25, -0) {};
		\node [style=none] (16) at (1.25, -0.4999999) {$\mathcal{S}_\phi$};
		\node [style=none] (17) at (0, 1) {};
		\node [style={right label}] (18) at (0, 0.5) {$A$};
		\node [style=none] (19) at (1.25, -0) {};
		\node [style=none] (20) at (0, -0) {};
		\node [style=none] (21) at (2.5, 1) {};
		\node [style=none] (22) at (2.5, -0) {};
		\node [style={right label}] (23) at (2.5, 0.5) {$P$};
		\node [style=upground] (24) at (2.5, 1.25) {};
	\end{pgfonlayer}
	\begin{pgfonlayer}{edgelayer}
		\draw (5.center) to (3.center);
		\draw (6.center) to (4.center);
		\draw (1.center) to (8.center);
		\draw (8.center) to (2.center);
		\draw (2.center) to (1.center);
		\draw (17.center) to (20.center);
		\draw (14.center) to (19.center);
		\draw (12.center) to (11.center);
		\draw (11.center) to (15.center);
		\draw (15.center) to (12.center);
		\draw (21.center) to (22.center);
	\end{pgfonlayer}
\end{tikzpicture}}$
\item $%
\begin{tikzpicture}
	\begin{pgfonlayer}{nodelayer}
		\node [style=none] (0) at (4, -0) {$=$};
		\node [style=none] (1) at (1.25, -0.9999999) {};
		\node [style=none] (2) at (-0.7500001, -0) {};
		\node [style={right label}] (3) at (1.25, 0.5) {$A$};
		\node [style=none] (4) at (1.25, 0.9999999) {};
		\node [style=none] (5) at (3.25, -0) {};
		\node [style=none] (6) at (1.25, -0.4999999) {$\mathcal{S}_\phi$};
		\node [style=none] (7) at (0, 1) {};
		\node [style={right label}] (8) at (0, 0.5) {$A$};
		\node [style=none] (9) at (1.25, -0) {};
		\node [style=none] (10) at (0, -0) {};
		\node [style=none] (11) at (2.5, 0.9999999) {};
		\node [style=none] (12) at (2.5, -0) {};
		\node [style={right label}] (13) at (2.5, 0.5) {$P$};
		\node [style=upground] (14) at (2.5, 1.25) {};
		\node [style=upground] (15) at (1.25, 1.25) {};
		\node [style=point] (16) at (5.25, -0.375) {$\mu_q$};
		\node [style=none] (17) at (5.25, 0.7500001) {};
	\end{pgfonlayer}
	\begin{pgfonlayer}{edgelayer}
		\draw (7.center) to (10.center);
		\draw (4.center) to (9.center);
		\draw (2.center) to (1.center);
		\draw (1.center) to (5.center);
		\draw (5.center) to (2.center);
		\draw (11.center) to (12.center);
		\draw (17.center) to (16);
	\end{pgfonlayer}
\end{tikzpicture}}$
\item $%
\begin{tikzpicture}
	\begin{pgfonlayer}{nodelayer}
		\node [style=none] (0) at (4, -0) {$=$};
		\node [style=none] (1) at (1.25, -0.9999999) {};
		\node [style=none] (2) at (-0.7500001, -0) {};
		\node [style=right label] (3) at (1.25, 0.5) {$A$};
		\node [style=none] (4) at (1.25, 0.9999999) {};
		\node [style=none] (5) at (3.25, -0) {};
		\node [style=none] (6) at (1.25, -0.4999999) {$\mathcal{S}_\phi$};
		\node [style=none] (7) at (0, 0.9999999) {};
		\node [style=right label] (8) at (0, 0.5) {$A$};
		\node [style=none] (9) at (1.25, -0) {};
		\node [style=none] (10) at (0, -0) {};
		\node [style=none] (11) at (2.5, 0.9999999) {};
		\node [style=none] (12) at (2.5, -0) {};
		\node [style=right label] (13) at (2.5, 0.5) {$P$};
		\node [style=upground] (14) at (2.5, 1.25) {};
		\node [style=copoint] (15) at (1.25, 1.25) {$0_q$};
		\node [style=point] (16) at (6, -0.25) {$\phi$};
		\node [style=none] (17) at (6, 0.75) {};
		\node [style=none] (18) at (5, -0) {$\frac{1}{d}$};
	\end{pgfonlayer}
	\begin{pgfonlayer}{edgelayer}
		\draw (7.center) to (10.center);
		\draw (4.center) to (9.center);
		\draw (2.center) to (1.center);
		\draw (1.center) to (5.center);
		\draw (5.center) to (2.center);
		\draw (11.center) to (12.center);
		\draw (17.center) to (16);
	\end{pgfonlayer}
\end{tikzpicture}}$
\een
Where the effect $0_q$ is the quantum effect $\mathrm{Tr}(\ketbra{0}{0}\_)$ which gives probability $1$ for state $0_q$ and probability $0$ for $1_q$.

{\color{black}
We will denote the Bell state $\frac{1}{d}\sum_{ij}\ketbra{ii}{jj}$ for a $d$-dimensional system diagrammatically as:
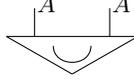
\[%
\begin{tikzpicture}
	\begin{pgfonlayer}{nodelayer}
		\node [style=none] (0) at (0, -0) {};
		\node [style=none] (1) at (0.7500001, -0) {};
		\node [style=none] (2) at (2.75, -0) {};
		\node [style=none] (3) at (3.5, -0) {};
		\node [style=none] (4) at (1.25, -0.25) {};
		\node [style=none] (5) at (2.25, -0.25) {};
		\node [style=none] (6) at (1.75, -1) {};
		\node [style=none] (7) at (0.75, 0.75) {};
		\node [style=none] (8) at (2.75, 0.75) {};
		\node [style=right label] (9) at (0.75, 0.75) {$A$};
		\node [style=right label] (10) at (2.75, 0.75) {$A$};
	\end{pgfonlayer}
	\begin{pgfonlayer}{edgelayer}
		\draw (0.center) to (6.center);
		\draw (6.center) to (3.center);
		\draw (3.center) to (0.center);
		\draw [bend right=90, looseness=1.50] (4.center) to (5.center);
		\draw (7.center) to (1.center);
		\draw (8.center) to (2.center);
	\end{pgfonlayer}
\end{tikzpicture}}\]
Recall that this has the maximally mixed state as it's marginals, that is:
\beq%
\begin{tikzpicture}
	\begin{pgfonlayer}{nodelayer}
		\node [style=none] (0) at (0, -0) {};
		\node [style=none] (1) at (0.7500001, -0) {};
		\node [style=none] (2) at (2.75, -0) {};
		\node [style=none] (3) at (3.5, -0) {};
		\node [style=none] (4) at (1.25, -0.25) {};
		\node [style=none] (5) at (2.25, -0.25) {};
		\node [style=none] (6) at (1.75, -1) {};
		\node [style=none] (7) at (0.7500001, 0.5000001) {};
		\node [style=none] (8) at (2.75, 0.75) {};
		\node [style=none] (9) at (7.75, 0.7499999) {};
		\node [style=none] (10) at (8.25, -0.25) {};
		\node [style=none] (11) at (8.75, -1) {};
		\node [style=none] (12) at (9.25, -0.25) {};
		\node [style=none] (13) at (7.75, -0) {};
		\node [style=none] (14) at (9.75, 0.5000001) {};
		\node [style=none] (15) at (7, -0) {};
		\node [style=none] (16) at (9.75, -0) {};
		\node [style=none] (17) at (10.5, -0) {};
		\node [style=none] (18) at (4, -0) {$=$};
		\node [style=point] (19) at (5.25, -0.25) {$\mu_q$};
		\node [style=none] (20) at (5.25, 0.75) {};
		\node [style=none] (21) at (6.5, -0) {$=$};
		\node [style=upground] (22) at (0.7500001, 0.75) {};
		\node [style=upground] (23) at (9.75, 0.75) {};
	\end{pgfonlayer}
	\begin{pgfonlayer}{edgelayer}
		\draw (0.center) to (6.center);
		\draw (6.center) to (3.center);
		\draw (3.center) to (0.center);
		\draw [bend right=90, looseness=1.50] (4.center) to (5.center);
		\draw (7.center) to (1.center);
		\draw (8.center) to (2.center);
		\draw (15.center) to (11.center);
		\draw (11.center) to (17.center);
		\draw (17.center) to (15.center);
		\draw [bend right=90, looseness=1.50] (10.center) to (12.center);
		\draw (9.center) to (13.center);
		\draw (14.center) to (16.center);
		\draw (20.center) to (19);
	\end{pgfonlayer}
\end{tikzpicture}}\label{eq:BellTrace1}\eeq
}
Then, as the parallel composition, i.e.~tensor product, of two pure quantum states is a pure quantum state, and the definition of hyperdecoherence {\color{black} ensures pure quantum states are pure post-quantum states} (point $3.$ of def.~(\ref{def:PQT})), the following is another purification of $\mu_q$ with the same purifying system of type $AP$ as $\mathcal{S}_\phi$
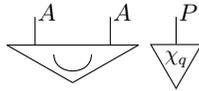
\[%
\begin{tikzpicture}
	\begin{pgfonlayer}{nodelayer}
		\node [style=none] (0) at (0, -0) {};
		\node [style=none] (1) at (0.7500001, -0) {};
		\node [style=none] (2) at (2.75, -0) {};
		\node [style=none] (3) at (3.5, -0) {};
		\node [style=none] (4) at (1.25, -0.25) {};
		\node [style=none] (5) at (2.25, -0.25) {};
		\node [style=none] (6) at (1.75, -1) {};
		\node [style=none] (7) at (0.7500001, 0.75) {};
		\node [style=none] (8) at (2.75, 0.75) {};
		\node [style=right label] (9) at (0.75, 0.75) {$A$};
		\node [style=right label] (10) at (2.75, 0.75) {$A$};
		\node [style=right label] (11) at (4.5, 0.75) {$P$};
		\node [style=none] (12) at (4.5, 0.7500001) {};
		\node [style=point] (13) at (4.5, -0.375) {$\chi_q$};
	\end{pgfonlayer}
	\begin{pgfonlayer}{edgelayer}
		\draw (0.center) to (6.center);
		\draw (6.center) to (3.center);
		\draw (3.center) to (0.center);
		\draw [bend right=90, looseness=1.50] (4.center) to (5.center);
		\draw (7.center) to (1.center);
		\draw (8.center) to (2.center);
		\draw (12.center) to (13);
	\end{pgfonlayer}
\end{tikzpicture}}\]
where $\chi_q$ is a pure quantum state. The purification principle implies that these two purifications are connected by a reversible transformation $R_\phi$:
\[%
\begin{tikzpicture}
	\begin{pgfonlayer}{nodelayer}
		\node [style=none] (0) at (0, -0) {};
		\node [style=none] (1) at (0.7500001, -0) {};
		\node [style=none] (2) at (2.75, -0) {};
		\node [style=none] (3) at (3.5, -0) {};
		\node [style=none] (4) at (1.25, -0.25) {};
		\node [style=none] (5) at (2.25, -0.25) {};
		\node [style=none] (6) at (1.75, -1) {};
		\node [style=none] (7) at (0.7500001, 1.5) {};
		\node [style=none] (8) at (2.75, 0.25) {};
		\node [style=none] (9) at (4.25, 0.25) {};
		\node [style=point] (10) at (4.25, -0.375) {$\chi_q$};
		\node [style=none] (11) at (3.5, 0.7500001) {$R_\phi$};
		\node [style=none] (12) at (2.5, 0.25) {};
		\node [style=none] (13) at (4.5, 0.25) {};
		\node [style=none] (14) at (4.5, 1.25) {};
		\node [style=none] (15) at (2.5, 1.25) {};
		\node [style=none] (16) at (2.75, 1.25) {};
		\node [style=none] (17) at (4.25, 1.25) {};
		\node [style=none] (18) at (4.25, 1.5) {};
		\node [style=none] (19) at (2.75, 1.5) {};
		\node [style=none] (20) at (5.75, -0) {$=$};
		\node [style=none] (21) at (6.5, -0) {};
		\node [style=none] (22) at (10.5, -0) {};
		\node [style=none] (23) at (8.5, -1) {};
		\node [style=none] (24) at (8.5, -0.5000001) {$\mathcal{S}_\phi$};
		\node [style=none] (25) at (7.25, -0) {};
		\node [style=none] (26) at (8.5, -0) {};
		\node [style=none] (27) at (9.75, -0) {};
		\node [style=none] (28) at (9.75, 0.7500001) {};
		\node [style=none] (29) at (8.5, 0.7500001) {};
		\node [style=none] (30) at (7.25, 0.7500001) {};
	\end{pgfonlayer}
	\begin{pgfonlayer}{edgelayer}
		\draw (0.center) to (6.center);
		\draw (6.center) to (3.center);
		\draw (3.center) to (0.center);
		\draw [bend right=90, looseness=1.50] (4.center) to (5.center);
		\draw (7.center) to (1.center);
		\draw (8.center) to (2.center);
		\draw (9.center) to (10);
		\draw (15.center) to (14.center);
		\draw (14.center) to (13.center);
		\draw (13.center) to (12.center);
		\draw (12.center) to (15.center);
		\draw (19.center) to (16.center);
		\draw (18.center) to (17.center);
		\draw (21.center) to (22.center);
		\draw (22.center) to (23.center);
		\draw (23.center) to (21.center);
		\draw (30.center) to (25.center);
		\draw (29.center) to (26.center);
		\draw (28.center) to (27.center);
	\end{pgfonlayer}
\end{tikzpicture}}\]

Using point $3.$ above, it then follows that there is an effect $e_\phi$ defined as: {\color{black}
\[
\begin{tikzpicture}
	\begin{pgfonlayer}{nodelayer}
		\node [style=none] (0) at (0, -0) {};
		\node [style=none] (1) at (0, -0.5) {};
		\node [style=copoint] (2) at (0, 0.25) {$e_\phi$};
	\end{pgfonlayer}
	\begin{pgfonlayer}{edgelayer}
		\draw (0.center) to (1.center);
	\end{pgfonlayer}
\end{tikzpicture}
:=
\begin{tikzpicture}
	\begin{pgfonlayer}{nodelayer}
		\node [style=none] (0) at (-0.75, -1.5) {};
		\node [style=none] (1) at (-0.75, -0.5) {};
		\node [style=none] (2) at (0.75, -0.5) {};
		\node [style=point] (3) at (0.75, -1) {$\chi_q$};
		\node [style=none] (4) at (0, -0) {$R_\phi$};
		\node [style=none] (5) at (-1, -0.5) {};
		\node [style=none] (6) at (1, -0.5) {};
		\node [style=none] (7) at (1, 0.5) {};
		\node [style=none] (8) at (-1, 0.5) {};
		\node [style=none] (9) at (-0.75, 0.5) {};
		\node [style=none] (10) at (0.75, 0.5) {};
		\node [style=none] (11) at (0.75, 0.75) {};
		\node [style=none] (12) at (-0.75, 0.75) {};
		\node [style=copoint] (13) at (-0.75, 1) {$0_q$};
		\node [style=upground] (14) at (0.75, 1) {};
	\end{pgfonlayer}
	\begin{pgfonlayer}{edgelayer}
		\draw (1.center) to (0.center);
		\draw (2.center) to (3);
		\draw (8.center) to (7.center);
		\draw (7.center) to (6.center);
		\draw (6.center) to (5.center);
		\draw (5.center) to (8.center);
		\draw (12.center) to (9.center);
		\draw (11.center) to (10.center);
	\end{pgfonlayer}
\end{tikzpicture}
\]
which} \emph{steers} the Bell state to $\phi$
\beq \label{steering1}
\begin{tikzpicture}
	\begin{pgfonlayer}{nodelayer}
		\node [style=none] (0) at (0, -0) {};
		\node [style=none] (1) at (0.7500001, -0) {};
		\node [style=none] (2) at (2.75, -0) {};
		\node [style=none] (3) at (3.5, -0) {};
		\node [style=none] (4) at (1.25, -0.25) {};
		\node [style=none] (5) at (2.25, -0.25) {};
		\node [style=none] (6) at (1.75, -1) {};
		\node [style=none] (7) at (0.7499999, 1.25) {};
		\node [style=none] (8) at (2.75, 0.9999999) {};
		\node [style=none] (9) at (4.25, 0.9999999) {};
		\node [style=point] (10) at (4.25, 0.5000002) {$\chi_q$};
		\node [style=none] (11) at (3.5, 1.5) {$R_\phi$};
		\node [style=none] (12) at (2.5, 0.9999999) {};
		\node [style=none] (13) at (4.5, 0.9999999) {};
		\node [style=none] (14) at (4.5, 2) {};
		\node [style=none] (15) at (2.5, 2) {};
		\node [style=none] (16) at (2.75, 2) {};
		\node [style=none] (17) at (4.25, 2) {};
		\node [style=none] (18) at (4.25, 2.25) {};
		\node [style=none] (19) at (2.75, 2.25) {};
		\node [style=none] (20) at (-2.25, 0.5000001) {};
		\node [style=none] (21) at (-5, -0) {};
		\node [style=none] (22) at (-2.75, -0.2500001) {};
		\node [style=none] (23) at (-4.25, -0) {};
		\node [style=none] (24) at (-3.25, -0.9999999) {};
		\node [style=none] (25) at (-1.5, -0) {};
		\node [style=none] (26) at (-2.25, -0) {};
		\node [style=none] (27) at (-4.25, 0.7499999) {};
		\node [style=none] (28) at (-3.75, -0.2500001) {};
		\node [style=none] (29) at (5.5, -0) {$=$};
		\node [style=none] (30) at (-0.7500001, -0) {$=$};
		\node [style=copoint] (31) at (-2.25, 0.7499999) {$e_\phi$};
		\node [style=copoint] (32) at (2.75, 2.5) {$0_q$};
		\node [style=upground] (33) at (4.25, 2.5) {};
		\node [style=point] (34) at (7.25, -0.25) {$\phi$};
		\node [style=none] (35) at (7.25, 0.75) {};
		\node [style=none] (36) at (6.25, -0) {$\frac{1}{d}$};
	\end{pgfonlayer}
	\begin{pgfonlayer}{edgelayer}
		\draw (0.center) to (6.center);
		\draw (6.center) to (3.center);
		\draw (3.center) to (0.center);
		\draw [bend right=90, looseness=1.50] (4.center) to (5.center);
		\draw (7.center) to (1.center);
		\draw (8.center) to (2.center);
		\draw (9.center) to (10);
		\draw (15.center) to (14.center);
		\draw (14.center) to (13.center);
		\draw (13.center) to (12.center);
		\draw (12.center) to (15.center);
		\draw (19.center) to (16.center);
		\draw (18.center) to (17.center);
		\draw (21.center) to (24.center);
		\draw (24.center) to (25.center);
		\draw (25.center) to (21.center);
		\draw [bend right=90, looseness=1.50] (28.center) to (22.center);
		\draw (27.center) to (23.center);
		\draw (20.center) to (26.center);
		\draw (35.center) to (34);
	\end{pgfonlayer}
\end{tikzpicture}}\eeq
{\color{black} Hence, for every pure state $\phi$ in the theory, there exists an effect, denoted $e_\phi$ that steers to it:
\beq \label{eq:steeringDiagram}
\begin{tikzpicture}
	\begin{pgfonlayer}{nodelayer}
		\node [style=none] (0) at (0.25, -0) {$d$};
		\node [style=none] (1) at (3.75, 0.5) {};
		\node [style=none] (2) at (1, -0) {};
		\node [style=none] (3) at (3.25, -0.25) {};
		\node [style=none] (4) at (1.75, -0) {};
		\node [style=none] (5) at (2.75, -1) {};
		\node [style=none] (6) at (4.5, -0) {};
		\node [style=none] (7) at (3.75, -0) {};
		\node [style=none] (8) at (1.75, 0.75) {};
		\node [style=none] (9) at (2.25, -0.25) {};
		\node [style=none] (10) at (-0.7500001, -0) {$=$};
		\node [style=copoint] (11) at (3.75, 0.75) {$e_\phi$};
		\node [style=point] (12) at (-2, -0.25) {$\phi$};
		\node [style=none] (13) at (-2, 0.75) {};
	\end{pgfonlayer}
	\begin{pgfonlayer}{edgelayer}
		\draw (2.center) to (5.center);
		\draw (5.center) to (6.center);
		\draw (6.center) to (2.center);
		\draw [bend right=90, looseness=1.50] (9.center) to (3.center);
		\draw (8.center) to (4.center);
		\draw (1.center) to (7.center);
		\draw (13.center) to (12);
	\end{pgfonlayer}
\end{tikzpicture}
\eeq
}

Using this steering result (Eq.~\ref{eq:steeringDiagram}) as well as Eq.~(\ref{localmap1})
and noting that the Bell state for a composite system is the composite of the Bell states for the single systems
\beq%
\begin{tikzpicture}
	\begin{pgfonlayer}{nodelayer}
		\node [style=none] (0) at (0, -0) {};
		\node [style=none] (1) at (0.7500001, -0) {};
		\node [style=none] (2) at (2.75, -0) {};
		\node [style=none] (3) at (3.5, -0) {};
		\node [style=none] (4) at (1.25, -0.25) {};
		\node [style=none] (5) at (2.25, -0.25) {};
		\node [style=none] (6) at (1.75, -1) {};
		\node [style=none] (7) at (0.7500001, 0.75) {};
		\node [style=none] (8) at (2.75, -0) {};
		\node [style=none] (9) at (4.75, -0) {};
		\node [style=none] (10) at (5.25, -0.25) {};
		\node [style=none] (11) at (5.75, -1) {};
		\node [style=none] (12) at (6.25, -0.25) {};
		\node [style=none] (13) at (4.75, -0) {};
		\node [style=none] (14) at (6.75, 0.7499999) {};
		\node [style=none] (15) at (4, -0) {};
		\node [style=none] (16) at (6.75, -0) {};
		\node [style=none] (17) at (7.5, -0) {};
		\node [style=right label] (18) at (0.75, 1.25) {$A$};
		\node [style=right label] (19) at (5.75, 1) {$A$};
		\node [style=right label] (20) at (1.75, 1.25) {$B$};
		\node [style=right label] (21) at (6.75, 1) {$B$};
		\node [style=none] (22) at (0.7500001, 1.25) {};
		\node [style=none] (23) at (1.75, 1.25) {};
		\node [style=none] (24) at (5.75, 1) {};
		\node [style=none] (25) at (6.75, 1) {};
		\node [style=none] (26) at (-0.75, -0) {$:=$};
		\node [style=none] (27) at (-4.25, -0) {};
		\node [style=right label] (28) at (-4.25, 1) {$AB$};
		\node [style=none] (29) at (-4.25, -0) {};
		\node [style=none] (30) at (-3.25, -1) {};
		\node [style=none] (31) at (-5, -0) {};
		\node [style=none] (32) at (-2.75, -0.25) {};
		\node [style=none] (33) at (-3.75, -0.25) {};
		\node [style=none] (34) at (-2.25, 1) {};
		\node [style=right label] (35) at (-2.25, 1) {$AB$};
		\node [style=none] (36) at (-3.25, 1) {};
		\node [style=none] (37) at (-1.5, -0) {};
		\node [style=none] (38) at (-2.25, 0.75) {};
		\node [style=none] (39) at (-4.25, 1) {};
		\node [style=none] (40) at (-2.25, -0) {};
	\end{pgfonlayer}
	\begin{pgfonlayer}{edgelayer}
		\draw (0.center) to (6.center);
		\draw (6.center) to (3.center);
		\draw (3.center) to (0.center);
		\draw [bend right=90, looseness=1.50] (4.center) to (5.center);
		\draw (7.center) to (1.center);
		\draw (8.center) to (2.center);
		\draw (15.center) to (11.center);
		\draw (11.center) to (17.center);
		\draw (17.center) to (15.center);
		\draw [bend right=90, looseness=1.50] (10.center) to (12.center);
		\draw (9.center) to (13.center);
		\draw (14.center) to (16.center);
		\draw (22.center) to (7.center);
		\draw [in=90, out=-90, looseness=0.50] (23.center) to (9.center);
		\draw [in=-90, out=90, looseness=0.50] (8.center) to (24.center);
		\draw (14.center) to (25.center);
		\draw (31.center) to (30.center);
		\draw (30.center) to (37.center);
		\draw (37.center) to (31.center);
		\draw [bend right=90, looseness=1.50] (33.center) to (32.center);
		\draw (29.center) to (27.center);
		\draw (38.center) to (40.center);
		\draw [in=90, out=-90, looseness=0.50] (39.center) to (29.center);
		\draw (38.center) to (34.center);
	\end{pgfonlayer}
\end{tikzpicture}}\eeq
we have, for all pure states $\psi$ and all effects $\eta$, that
\[%
\InputIfFileExists{Diagrams/lastStep.tikz}{}{\input{./figures/Diagrams/lastStep.tikz}}\]
This result, in conjunction with tomography (Eq.~(\ref{tomography})) and convexity (Eq.~(\ref{convexity})), implies that, for all system types $A$,\vspace{-.25cm}
\[%
\begin{tikzpicture}
	\begin{pgfonlayer}{nodelayer}
		\node [style=none] (0) at (0, -0) {};
		\node [style=none] (1) at (0, -1.5) {};
		\node [style=proj] (2) at (0, -0.75) {};
		\node [style=none] (3) at (1.5, -0.75) {$=$};
		\node [style=none] (4) at (2.5, -0) {};
		\node [style=none] (5) at (2.5, -1.5) {};
		\node [style={right label}] (6) at (0.25, -1) {$A$};
		\node [style={right label}] (7) at (2.5, -1) {$A$};
	\end{pgfonlayer}
	\begin{pgfonlayer}{edgelayer}
		\draw (0.center) to (2);
		\draw (2) to (1.center);
		\draw (4.center) to (5.center);
	\end{pgfonlayer}
\end{tikzpicture}}\]
\end{proof}

As we know that there exists a post-classical theory which satisfies causality and purification and decoheres to classical theory, i.e. quantum theory, one might wonder at what stage our proof breaks down when analysing this situation. The main reason is that the maximally correlated state in classical probability theory is mixed and so {\color{black}the classical analogue to the state \ref{eq:newPurification} is not a purification and} Eq.~(\ref{steering}) is no longer valid. Hence, the reason why quantum theory cannot be extended in the manner proposed here is the existence of pure entangled states.

\section{Discussion}\label{sec:Discussion}

From the famous theorems of Bell \cite{bell1964einstein} and Kochen \& Specker \cite{kochen1967problem} to more recent results by Colbeck \& Renner \cite{colbeck2011no}, and Pusey, Barrett \& Rudolph \cite{pusey2012reality}, no-go theorems have a long history in the foundations of quantum theory. Most previous no-go theorems have been concerned with ruling out certain classes of hidden variable models from some set of natural assumptions. Hidden variables---or their contemporary incarnation as ontological models \cite{harrigan2010einstein}---aim to provide quantum theory with an underlying classical description, where non-classical quantum features arise due to the fact that this description is `hidden' from us.

Unlike these approaches, our result rules out certain classes of operationally-defined physical theories which can supersede quantum theory, yet reduce to it via a suitable process. To the best of our knowledge, our no-go theorem is the first of its kind. This may seem surprising given that it is an obvious question to ask. However, to even begin posing such questions in a rigorous manner requires a consistent way to define operational theories beyond quantum and classical theory. The mathematical underpinnings of such a framework have only recently been developed and investigated in the field of quantum foundations.

As with all no-go theorems, our result is only as strong as the assumptions which underlie it. We now critically examine each of our assumptions, outlining for each one the sense in which it can be considered `natural', yet also suggesting ways in which a hypothetical post-quantum theory could violate it and hence escape the conclusion of our theorem.

Our first assumption is purification. As noted in section~\ref{Generalised theories}, the purification principle provides a way of formalising the natural idea that information can only be discarded \cite{chiribella2015conservation}, and any lack of information about the state of a given system arises in an essentially unique way due to a lack of information about some larger environment system. However, proposals for constructing theories in which information can be fundamentally destroyed have been suggested and investigated \cite{oppenheim2009fundamental, banks1984difficulties, unruh1995evolution}. Such proposals take their inspiration from the Black Hole Information Loss paradox. Our result can therefore be thought of as providing another manner in which the fundamental status of information conservation can be challenged.

Our second assumption is causality. This principle allows one to uniquely define a notion of ``past'' and ``future'' for a given process in a diagram, and is equivalent to the statement that future measurement choices do not affect current experimental outcomes. As such, this principle appears to be fundamental to the scientific method. Despite this, recent work has shown how one can relax this principle to arrive at a notion of ``indefinite'' causality \cite{oreshkov2012quantum, oreshkov2014operational, chiribella2016quantum, hardy2007towards}. In this case, there may be no matter of fact about whether a given process causally precedes another. The indefinite causal order between two processes has even been shown to be a resource which can be exploited to outperform theories satisfying the causality principle in certain information-theoretic tasks \cite{araujo2014computational, chiribella2012perfect}. Moreover, it has been suggested that any theory of Quantum Gravity must exhibit indefinite causal order \cite{hardy2016operational,hardy2016reconstructing}. Hence, as in the previous paragraph, our result provides further motivation for discarding the notion of definite causal order in the search for theories superseding quantum theory.

As purification seems to require a unique way to marginalise multipartite states, one might wonder whether one can define a notion of purification without the causality principle. {\color{black} Indeed, recent work \cite{chiribella2016quantum} has shown how one can formalise a purification principle in the absence of causality, \cite{Araujo2016purification} shows how an alternative notion of purification can be defined for process-matrices allowing for indefinite causal order, and work of one of the authors discusses a `time-symmetric' notion of purification satisfied by quantum, classical and hybrid quantum-classical systems \cite{selby2017classical}.}

Another assumption in our theorem was the manner in which our hyperdecoherence map---the mechanism by which the post-quantum theory reduces to quantum theory---was formalised. It may not be the case that post-quantum physics gives rise to quantum physics via such a mechanism. Indeed, alternate proposals for how some hypothetical post-quantum theory reduces to quantum theory have been proposed \cite{kleinmann2013typical}. Moreover, there is some evidence from research in quantum gravity that quantum pure states may become mixed at short length scales \cite{muller2009does}. This suggests that quantum pure states may not be fundamentally pure in a full theory of quantum gravity. However, we see the necessity of the requirement that quantum pure states are pure in a potential post-quantum theory (point $3.$ from def.~(\ref{def:PQT})) in our derivation as a feature rather than a bug. Indeed, it lends evidence to the assertion that to supersede quantum theory one must give up the requirement that states which appear pure within quantum theory are fundamentally pure. Despite this, our understanding of the quantum to classical transition in terms of decoherence suggests hyperdecoherence as the natural mechanism by which this should occur. Moreover, as discussed in section~\ref{sec:Hyperdecoherence} and shown in appendix~\ref{app:PureArePure}, one can derive that pure quantum states are pure post-quantum states from more primitive notions.

The last assumption underlying our no-go theorem is the generalised framework itself, introduced in section~\ref{Generalised theories}. \color{black} While the operational methodology and assumptions underlying this framework seem to be relatively mild,
it may not be the case that the correct way to formalise this methodology is by asserting that pieces of laboratory equipment can be composed together to result in experiments, as described in section~\ref{Generalised theories}. Indeed, it may be the case that the standard manner in which elements of a theory are composed together---resulting in other elements---needs to be revised in order to go beyond the quantum formalism. Work in this direction has already begun \cite{hardy2013theory}. Alternatively, one could take a more radical position and avoid this no-go result by accepting that a more fundamental theory of nature will not have an operational description at all, and that this framework and the operational methodology should be abandoned in their totality.\color{black}

Our result can either be viewed as demonstrating that the fundamental theory of Nature is quantum mechanical, or as showing in a rigorous manner that any post-quantum theory must radically depart from a quantum description of the world by abandoning the principle of causality, the principle of purification, or both.

\section*{Acknowledgements}
The authors thank H. Barnum and B. Coecke for discussions and D. Browne, M. Hoban, $\&$ J. Richens for proof reading a draft. They also acknowledge encouragement from J. J. Barry.
\section*{Funding statement}
This work was supported by the EPSRC through the UCL EPSRC Doctoral Prize Fellowship and the Controlled Quantum Dynamics Centre for Doctoral Training. This work began while the authors were attending ``Formulating and Finding Higher-order Interference'' at PI. Research at Perimeter Institute is supported by the Government of Canada through the Department of Innovation, Science and Economic Development Canada and by the Province of Ontario through the Ministry of Research, Innovation and Science.
\section*{Author contributions}
Both authors contributed equally to the current work.
\section*{Data accessibility}
This paper has no data.
\section*{Competing interests}
The authors declare no competing financial interests.

\bibliographystyle{plain}
\bibliography{library}

\appendix
\section{Algebraic proof of Main Theorem}\label{app:DiagrammaticProof}
We now present the proof of our Main Theorem using algebraic notation in place of the diagrammatic notation used in the proof presented in section~\ref{sec:Results}.

\begin{proof}
For convenience we denote quantum states with a superscript $q$. As discussed at the end of section~\ref{Generalised theories}, given a bipartite quantum state $\psi^q$, it can always be written as
\[
\psi^q_{AB}=\sum_{ij}r_{ij}\phi^q_{iA}\otimes \chi^q_{jB}, \qquad r_{ij}\in \mathds{R}.
\]
\color{black} Where the fact that this holds even when representing quantum theory as a sub-theory of the post-quantum theory follows immediately from, i) the definition of a sub-theory, and ii) linearity of transformations. \color{black} Idempotence of the hyperdecoherence map (item $2.$ from def.~(\ref{def:PQT})) then gives
\beq \label{localmap}
\mathds{1}_A\otimes \mathcal{D}_B [\psi^q_{AB}]=\sum_{ij}r_{ij}\phi^q_{iA}\otimes \mathcal{D}_B[\chi^q_{jB}]=\psi^q_{AB}
\eeq

Next, consider the maximally mixed quantum state, $$\mu^q:=\frac{\mathds{1}}{d},$$ of a $d$-dimensional system. {\color{black} By $3b.$ of def.~(\ref{def:PQT})) this is maximally mixed for the post-quantum theory, hence for any pure
state
$\psi$, there is a state $\sigma$
such that
\beq \label{eq:anyDecomposition}
\mu^q=\frac{1}{d} \psi +\left(1-\frac{1}{d}\right)\sigma
\eeq
}that is, \emph{any} pure state from the post-quantum theory arises in a decomposition of the quantum maximally mixed state.

Recall that every (non-trivial) quantum system of type $A$ has at least two perfectly distinguishable states, denoted here as $\{ \mathbf{0}^q:= \ketbra{0}{0}, \mathbf{1}^q:= \ketbra{1}{1}\}$. Given the decomposition of Eq.~(\ref{eq:anyDecomposition}), \emph{convexity} (Eq.~(\ref{convexity})) implies the following is a state in the post-quantum theory:
\[
s^\phi_{A_1A_2}:=\frac{1}{d}\phi_{A_1}\otimes \mathbf{0}_{A_2}^q+\left(1-\frac{1}{d}\right)\sigma_{A_1}\otimes \mathbf{1}_{A_2}^q
\]

Consider a purification of this state, denoted $\mathcal{S}^\phi_{A_1A_2P}$, and note that it has the following properties:
\ben
\item   $\mathsf{u}_P[\mathcal{S}^\phi_{A_1A_2P}]=s^\phi_{A_1A_2}$
\item   \color{black}$(\mathsf{u}_{A_2}\otimes \mathsf{u}_{P})[\mathcal{S}^\phi_{A_1A_2P}]=\mu^q_{A_1}$\color{black}
\item   $(e_{0A_2}^q\otimes\mathsf{u}_P)[\mathcal{S}^\phi_{A_1A_2P}]=\frac{1}{d}\phi_{A_1}$
\een
Where the effect  $e_0^q$ is the quantum effect $\mathrm{Tr}(\ketbra{0}{0}\_)$ which gives probability $1$ for state  $\mathbf{0_q}$ and probability $0$ for  $\mathbf{1}_q$ .

Now, let us denote the Bell state $\frac{1}{d}\sum_{ij}\ketbra{ii}{jj}$ for a $d$-dimensional system of type $A$ as:
$\mathcal{B}_{A_1A_2}^q, $
where $A_1$ and $A_2$ are the same type of system, but with a dummy index \color{black} to allow us keep track of their ordering algebraic notation \color{black}.
As the hyperdecoherence map is {\color{black} terminal} (point $1.$ from def.~(\ref{def:PQT})), marginalisation in the post-quantum theory is the same as in quantum theory. Hence, as shown in Eq.~\ref{marginalised Bell state 3} from section~\ref{Generalised theories}, \color{black} both of \color{black} the marginals of the above Bell state are equal to the maximally mixed quantum state:
\color{black}
\[
\begin{aligned}
\mathsf{u}_{A_1}[\mathcal{B}_{A_1A_2}^q]&=\mathrm{Tr}_{A_1}\left(\frac{1}{d}\sum_{ij}\ketbra{ii}{jj} \right)=\frac{\mathds{1}}{d},\\
\mathsf{u}_{A_2}[\mathcal{B}_{A_1A_2}^q]&=\mathrm{Tr}_{A_2}\left(\frac{1}{d}\sum_{ij}\ketbra{ii}{jj} \right)=\frac{\mathds{1}}{d}.
\end{aligned}
\]
As mentioned in section~\ref{Generalised theories}, the only relevant data regarding a process are the types and orderings of the inputs and outputs, where the ordering is kept track of with additional dummy index on the types when ambiguity could arise. In this case however, each of the resulting states has only a single output. Hence there is no ambiguity and we can drop this additional dummy index and write:
\beq\label{eq:BellTrace}\mathsf{u}_{A_1}[\mathcal{B}_{A_1A_2}^q]=\frac{\mathds{1}_{A_2}}{d}=\frac{\mathds{1}_{A}}{d}=\frac{\mathds{1}_{A_1}}{d}=\mathsf{u}_{A_2}[\mathcal{B}_{A_1A_2}^q]\eeq

 As the  parallel composition, i.e. tensor product, of two pure quantum states is a pure quantum state, and the definition of hyperdecoherence ensures pure quantum states are pure post-quantum states (point $3.$ of def.~(\ref{def:PQT})), the following is another purification of  $\mu^q$ with the same purifying system of type $AP$ as  $\mathcal{S}^\phi$
\beq \label{eq:newPurification}
\mathcal{B}^q_{A_1A_2}\otimes \chi^q_P
\eeq
where  $\chi^q$ is a pure quantum state. The purification principle implies that these two purifications are connected by a reversible transformation  $R^\phi_{A_2P}$:
\[
\mathds{1}_{A_1}\otimes R^\phi_{A_2P}[\mathcal{B}^q_{A_1A_2}\otimes \chi^q_P]=\mathcal{S}^\phi_{A_1A_2P}
\]

Using point $3.$ above, it then follows that there is an effect  $ e_{\phi }^A$, {\color{black} defined as:
\[
e_{\phi}^{A_2}[\ \_\ ]:= (e_{0}^{A_2 q}\otimes \mathsf{u}_P)[R^\phi_{A_2P}[\ \_\ \otimes \chi^q_P]]
\]
which} \emph{steers} the Bell state to  $\phi_A$, {\color{black} that is}:
\[
\begin{aligned}
e_{\phi}^{A_2}[\mathcal{B}^q_{A_1A_2}]=& (e_{0}^{A_2 q}\otimes \mathsf{u}_P)[\mathds{1}_{A_1}\otimes R^\phi_{A_2P}[\mathcal{B}^q_{A_1A_2}\otimes \chi^q_P]] \\
&=\frac{1}{d}\phi_{A_1}
\end{aligned}
\]
Hence for every pure state $\phi$ in the theory, there exists an effect, denoted $e_\phi$, that steers to it:
\beq \label{steering}
\phi_{A_1}=d_{A_1}\left( e_{\phi}^{A_2}\circ\left[\mathcal{B}^q_{A_1A_2}\right]\right),
\eeq
where $d_{A_1}$ is the dimension of the system of type $A_1$.

\color{black}
Using this steering result (Eq.~(\ref{steering})) as well as Eq.~(\ref{localmap}) we can immediately see that the hyperdecoherence map must act as the identity on all states:
\[\begin{array}{l}
\mathcal{D}_{A_1}\circ\phi_{A_1}\\ \quad \stackrel{\ref{steering}}{=}d_{A_1}\left( \mathcal{D}_{A_1}\otimes e_{\phi}^{A_2}\circ\left[\mathcal{B}^q_{A_1A_2}\right]\right)\\ \quad \stackrel{\ref{localmap}}{=}d_{A_1}\left( \mathds{1}_{A_1}\otimes e_{\phi}^{A_2}\circ\left[\mathcal{B}^q_{A_1A_2}\right]\right)\\ \quad \stackrel{\ref{steering}}{=}\phi_{A_1}
\end{array} \]\color{black}
where the overline numbers refer to the equations used to obtain each equality. \color{black} This can easily be extended to the case when $\mathcal{D}$ is acting on an arbitrary (and possibly composite) system of a composite state, by using the steering result for a Bell state of a composite system.\color{black}
This result, in conjunction with tomography (Eq.~(\ref{tomography})) and convexity (Eq.~(\ref{convexity})) implies that, for all systems of type $A$
\[
\mathcal{D}_A=\mathds{1}_A
\]
\end{proof}

\vspace{-.5cm}
\section{Proof that pure quantum states are pure} \label{app:PureArePure}\label{Proof that pure quantum states are pure}

In section~\ref{sec:Hyperdecoherence}
we discussed how one of the key features of quantum to classical decoherence was that pure classical states are also pure when considered within quantum theory.  We took a generalisation of this as a defining feature of hyperdecoherence to prove our main theorem. In particular we noted how this seemed essential to ruling out particular cases, such as a bit ``decohering'' from a pair of bits, which satisfy {\color{black} terminality} and idempotence, but fail to adequately capture the spirit of decoherence. However, as noted previously, these examples are also ruled out by a seemingly weaker condition; (hyper)decoherences preserves the \emph{information dimension} of the system. Before we present the definition of information dimension, recall that two states $\rho_1$ and $\rho_2$ are perfectly distinguishable if there exists a measurement $\{e_1, e_2 \}$ such that $e_i\left[ \rho_j \right]=\delta_{ij}$.

\begin{definition}[Information dimension \cite{brunner2014dimension}] \label{information dimension}
The information dimension of a system is the number of states in a maximal set that are all pairwise perfectly distinguishable.
\end{definition}
Note that for a quantum or classical $d$-level system, the information dimension is $d$ \cite{brunner2014dimension}, hence standard quantum to classical decoherence preserves the information dimension. However, in the example presented in section~\ref{sec:Hyperdecoherence},
 this is not the case. For instance, if two bits ``decoheres'' to a bit, then the information dimension goes from $4$ to $2$. Hence, in place of point $3.$ from def.~(\ref{def:PQT}), we could have stipulated that the hyperdecoherence map \emph{preserves information dimension}.

We will now show that from i) preservation of information dimension and ii) a common strengthening of the notion of purification \cite{chiribella2010probabilistic}, one can derive the previously postulated requirement that pure quantum states be pure in the post-quantum theory.

\begin{definition}[Strong purification] \label{strong purification}
\
\begin{enumerate}
\item Every mixed state of system of type $A$ can be purified to a state of system of type $AA$ satisfying def.~(\ref{def:purification}).
\item If a state $\rho$ of system of type $A$ is pure, then it has trivial purifications on all systems. That is, it has a purification $\psi$ on system of type $AB$ which factorises as $\psi = \rho \otimes \chi$, where $\chi$ is a state of $B$, for all system types $B$.
\end{enumerate}
\end{definition}
We now provide an outline of the proof before providing the formal argument below.

\paragraph*{Outline:}

Recall that every quantum pure state is an element of a maximal set of pairwise perfectly distinguishable quantum states. Assume toward contradiction that at least one quantum state is mixed in the post-quantum theory, and decompose it as a convex combination of post-quantum states. Every post-quantum state in this decomposition is perfectly distinguishable from any state the original quantum state is distinguishable from. Using strong purification, we show that there must be a pair of perfectly distinguishable post-quantum states in this decomposition. Hence, we have at least an information dimension of $d_Q+1$, where $d_Q$ is the quantum information dimension. Therefore, as we are assuming that information dimension is preserved, we must reject the assumption that any pure quantum state is mixed in the full theory.

\paragraph*{Proof:}
Consider a set of pure \& perfectly distinguishable quantum states $\{\textbf{i}^q := \ketbra{i}{i}\}_{i=0}^{d_Q-1}$ where $d_Q$ is the information dimension of the quantum system.
Recall that every quantum pure state is an element of a maximal set of pairwise perfectly distinguishable quantum states. Assume towards contradiction that at least one of the above set of quantum pure states is mixed in the full theory, without loss of generality we take this to be $\textbf{0}^q$. We can therefore write it as:
\[\textbf{0}^q=p s+(1-p)\sigma \]
where $0<p<1$. We pick a decomposition such that $p$ is maximised over all possible pure states $s$. Note, compactness of the set of states (which follows from purification \cite{chiribella2010probabilistic,chiribella2011informational}) ensures that such a maximum exists. In particular, the maximality of $p$ on $s$ means that
\beq\label{no decomp}
\sigma=q s+(1-q)\tau \text{ } \implies \text{ } q=0.
\eeq
Note that, due to purity of $\textbf{0}^q$ in quantum theory, $s$ and $\sigma$ must both hyperdecohere to $\textbf{0}^q$. That is, as
\[\textbf{0}^q=\mathcal{D}[\textbf{0}^q]=p \mathcal{D}[s]+(1-p)\mathcal{D}[\sigma] \]
is a pure quantum state and $\mathcal{D}[s]$ and $\mathcal{D}[\sigma]$ are quantum states, we must have
\beq \label{hype}
\mathcal{D}[s]=\mathcal{D}[\sigma] =\mathcal{D}[\mathbf{0}^q]=\mathbf{0}^q
\eeq
to ensure quantum purity of $\textbf{0}^q$.

To proceed, we need the following lemmas, which shall be proved later in this appendix. Before we state our first lemma, recall that the set of states appearing in the convex decomposition of a mixed state are said to \emph{refine} it.
\begin{lemma} \label{max. mixed} \
There exists a unique state, denoted $\mu_{PQ}$, of the full theory that is invariant under all reversible transformations and, moreover, is maximally mixed, i.e. it is \emph{refined} by any other state.
\end{lemma}
\begin{lemma} \label{pure effect}
For every pure state $a$ in the full theory, there exists a pure, i.e. unrefinable, effect $e$ on the same system such that $e[a]=1$.
\end{lemma}
\begin{lemma} \label{pure state on max. mixed}
Every pure effect gives the same probability on $\mu_{PQ}$, that is $p_*:=e[\mu_{PQ}] \text{, } \forall \text{ pure } e$.
\end{lemma}
We have the following consequence of the conjunction of transitivity and the above lemma, see section D of \cite{chiribella2016entanglement} for the proof.
\begin{corollary} \label{max. prob}
For any pure state $a$, the maximally mixed post-quantum state can be decomposed as \[\mu_{PQ}=p_* a+(1-p_*)\alpha\]
where $p_*$ is the maximal possible probability for any pure state in a decomposition of $\mu_{PQ}$.
\end{corollary}

Now, from {\color{black}point $3b.$ of def.~\ref{def:PQT}} and the form of the quantum maximally mixed state, we can write:
\[\mu_{PQ}=\mu_{Q}=\frac{1}{d_Q}\sum_{i=0}^{d_Q-1}\mathbf{i}^q\]
which, given our decomposition of $\mathbf{0}^q$ can be written:
\[\mu_{PQ}=\frac{p}{d_Q}s+ \left(\frac{1-p}{d_Q}\sigma+\frac{1}{d_Q}\sum_{i=1}^{d_Q-1}\mathbf{i}^q \right)\]
We know from Eq.~(\ref{no decomp}) that $s$ cannot appear in any decomposition of $\sigma$, and, moreover, $s$ cannot appear in any possible convex decomposition of the other $\mathbf{i}^q$. If it did, then, recalling Eq.~(\ref{hype}), $s$ would hyperdecohere to more than one quantum state, which is not possible as $\mathcal{D}$ is an idempotent linear map (point $2.$ of def.~(\ref{def:PQT})).
Hence, the maximal probability for $s$ appearing in a decomposition of $\mu_{PQ}$ (see corollary~\ref{max. prob}) is given by
\beq \label{maximality}
p_*=\frac{p}{d_Q}.
\eeq
Now consider the pure effect $e$ such that $e[s]=1$, which we know exists from lemma~\ref{pure effect}. We know, from lemma~\ref{pure state on max. mixed}, that $e[\mu_{PQ}]=p_*$. This, in conjunction with Eq.~(\ref{maximality}), gives:
\begin{align*}
p_* &=e[\mu_{PQ}]\\
&=\frac{p}{d_Q}e[s]+e\left[\frac{1-p}{d_Q}\sigma+\frac{1}{d_Q}\sum_{i=1}^{d_Q-1}\mathbf{i}^q \right]\\
&=p_*+e\left[\frac{1-p}{d_Q}\sigma+\frac{1}{d_Q}\sum_{i=1}^{d_Q-1}\mathbf{i}^q \right]
\end{align*}
and so,
\[e[\sigma]=0 \qquad \& \qquad e[\mathbf{i}^q]=0,\ \ \forall \textbf{i}^q \neq \textbf{0}^q\]
In particular this implies $s$ and $\sigma$ are perfectly distinguishable using the measurement $\{e, u-e\}$. This means that we can perfectly distinguish the two sets $\{s,\sigma\}$ and $\{\mathbf{i}^q\}_{i=1}^{d_Q-1}$ and moreover, the elements of each individual set are pairwise perfectly distinguishable. The information dimension (def.~(\ref{information dimension})) is therefore at least $d_Q+1$. Hence we must reject the assumption that pure quantum states are mixed in the larger theory.

\paragraph*{Proof of lemma~\ref{max. mixed}:} This has been shown to hold in any theory satisfying purification and convexity by \cite[Corollary $34$]{chiribella2010probabilistic}. \

\paragraph*{Proof of lemma~\ref{pure effect}:}
First, note that strong purification (def.~(\ref{strong purification})) implies that the parallel composition of pure states is a pure state. That is, if $a$ and $b$ are pure states, then strong purification implies $a\otimes b$ is also a pure state. Indeed point $2.$ in def.~(\ref{strong purification}) states that every pure state $\rho$ on any system of type $A$ has a trivial purification $\phi = \rho \otimes \chi$ on system of type $AB$, for all system types $B$. As $\phi$ is pure, $\chi$ must be as well, otherwise $\phi$ would have a non-trivial convex decomposition. As transitivity implies any two pure states are connected by a reversible transformation, and applying a reversible transformation to a pure state results in a pure state, we thus have that the parallel composition of any two pure states results in a pure state.

Now, consider the quantum state $\mathbf{0}^q$ and quantum effect $e^q_0$ which picks it out with probability $1$. Take any convex decomposition of this quantum state into post-quantum states $\{s, \sigma\}$, then:
\beq \label{picked out with probability 1}
\begin{tikzpicture}
	\begin{pgfonlayer}{nodelayer}
		\node [style=copoint] (0) at (-1.5, 0.75) {$e^q_0$};
		\node [style=point] (1) at (-1.5, -0.75) {$0^q$};
		\node [style=none] (2) at (-3.5, -0) {$1$};
		\node [style=none] (3) at (-2.75, -0) {$=$};
		\node [style=none] (4) at (1.25, -0.25) {$=\sum\limits_{x\in\{s, \sigma\}} p_x$};
		\node [style=point] (5) at (4, -0.75) {$x$};
		\node [style=copoint] (6) at (4, 0.75) {$e_0^q$};
		\node [style=none] (7) at (6, -0) {$\implies$};
		\node [style=point] (8) at (8, -0.75) {$x$};
		\node [style=copoint] (9) at (8, 0.75) {$e^q_0$};
		\node [style=none] (10) at (10, -0) {$1$};
		\node [style=none] (11) at (9.25, -0) {$=$};
	\end{pgfonlayer}
	\begin{pgfonlayer}{edgelayer}
		\draw (0) to (1);
		\draw (6) to (5);
		\draw (9) to (8);
	\end{pgfonlayer}
\end{tikzpicture}
\eeq

Now, consider the following convex combination of perfectly distinguishable pure states $\nu = p_0 s + \sum_{i=1}^{d_Q-1} p_i \textbf{i}^q$, where $p_i >0$ for all $i$, noting that perfect distinguishability of $s$ from the states $\{\textbf{i}^q\}_{i\neq 0}$ follows from Eq.~(\ref{picked out with probability 1}) and the fact that $\{\textbf{i}^q\}$ are perfectly distinguishable. If any of the $\textbf{i}^q$ are mixed, replace them with a pure post-quantum state that refines them to ensure $\nu$ is a convex combination of pure and perfectly distinguishable states. This state must be a maximally mixed state, i.e. every other state refines it, otherwise there would be a state that could be perfectly distinguished from it, violating preservation of information dimension.

Take a purification of $\nu$:
\[
\begin{tikzpicture}
	\begin{pgfonlayer}{nodelayer}
		\node [style=none] (0) at (0.5, 1.25) {};
		\node [style=none] (1) at (0.5, -0) {};
		\node [style=upground] (2) at (2, 1.25) {};
		\node [style=none] (3) at (2, -0) {};
		\node [style=point] (4) at (-3.5, -0.25) {$\nu$};
		\node [style=none] (5) at (-3.5, 1) {};
		\node [style=none] (6) at (-1.5, -0) {$=$};
		\node [style=none] (7) at (1.25, -0.5) {$\psi$};
		\node [style=none] (8) at (0, -0) {};
		\node [style=none] (9) at (1.25, -1) {};
		\node [style=none] (10) at (2.5, -0) {};
		\node [style=right label] (11) at (-3.5, 1) {$A$};
		\node [style=right label] (12) at (0.5, 1.25) {$A$};
		\node [style=right label] (13) at (2, 0.5) {$A$};
	\end{pgfonlayer}
	\begin{pgfonlayer}{edgelayer}
		\draw (0.center) to (1.center);
		\draw (2) to (3.center);
		\draw (5.center) to (4);
		\draw (8.center) to (10.center);
		\draw (10.center) to (9.center);
		\draw (9.center) to (8.center);
	\end{pgfonlayer}
\end{tikzpicture}\]
Corollary $12$ of \cite{chiribella2010probabilistic} states that if the marginal of a bipartite pure state on a given subsystem can be decomposed as a convex combination of perfectly distinguishable states, then the marginal on the opposite subsystem can also be decomposed as a convex combination of the same number of perfectly distinguishable states. As $\nu$ is a convex combination of pure and perfectly distinguishable states, we thus have:
\beq
\label{max. mixed marginal}
\begin{tikzpicture}
	\begin{pgfonlayer}{nodelayer}
		\node [style=none] (0) at (1.5, 1.5) {};
		\node [style=none] (1) at (1.5, -0) {};
		\node [style=upground] (2) at (0, 1.25) {};
		\node [style=none] (3) at (0, -0) {};
		\node [style=point] (4) at (6.25, -0.25) {$\rho_i$};
		\node [style=none] (5) at (6.25, 1) {};
		\node [style=none] (6) at (4, -0) {$=\sum_i p_i$};
		\node [style=none] (7) at (0.75, -0.5) {$\psi$};
		\node [style=none] (8) at (-0.5, -0) {};
		\node [style=none] (9) at (0.75, -1) {};
		\node [style=none] (10) at (2, -0) {};
		\node [style=right label] (11) at (6.25, 0.75) {$A$};
		\node [style=right label] (12) at (1.75, 1.25) {$A$};
		\node [style=right label] (13) at (0, 0.5) {$A$};
	\end{pgfonlayer}
	\begin{pgfonlayer}{edgelayer}
		\draw (0.center) to (1.center);
		\draw (2) to (3.center);
		\draw (5.center) to (4);
		\draw (8.center) to (10.center);
		\draw (10.center) to (9.center);
		\draw (9.center) to (8.center);
	\end{pgfonlayer}
\end{tikzpicture}
\eeq
where $i=0,...,d_Q-1$ and the $\rho_i$ are perfectly distinguishable. This marginal state must again be a maximally mixed state, i.e. every other state refines it, otherwise there would be a state that could be perfectly distinguished from it, violating preservation of information dimension.

It was proved in corollary $8$ of \cite{chiribella2010probabilistic} that in any theory satisfying purification and in which the parallel composition of pure states is pure, every bipartite pure state is \emph{steering} for its marginals. That is, any state which refines the marginal of a pure bipartite state can be \emph{steered} to by applying an effect to the opposite system. Moreover, corollary $12$ of \cite{chiribella2010probabilistic} ensures that the set of effects which, when applied to $\psi$, steer to $s$ and $\{\textbf{i}^q\}_{i\neq 0}$ correspond to the effects which perfectly distinguish among the states $\{\rho_i\}$. In particular, the effect $e_{\rho_0}$ which picks out $\rho_0$ is the effect that steers to $s$. Additionally, the effects which distinguish the states $s$ and $\{\textbf{i}^q\}_{i\neq 0}$ are the ones which steer to the states $\{\rho_i\}$ when applied to $\psi$.

To complete the proof we need one more ingredient: the States-Transformations isomorphism, or generalised Choi theorem \cite[Theorem 17]{chiribella2010probabilistic}. This theorem implies that if the marginals of a bipartite pure state are both maximally mixed, then any effect which steers to a pure state on either system must be pure. The generalised Choi theorem holds in any theory satisfying purifcation, in which the parallel composition of pure states is pure, and in which the product of maximally mixed states is maximally mixed (which holds for us due to 
point $3b.$ of def.~\ref{def:PQT}). In particular this implies that the effect $e_{\rho_0}$, which steers to $s$, must be pure.

Combining this with Eq.~(\ref{picked out with probability 1}), and the discussion after Eq.~(\ref{max. mixed marginal}) we have that:
\[
\begin{tikzpicture}
	\begin{pgfonlayer}{nodelayer}
		\node [style=copoint] (0) at (2.5, 0.75) {$e_0^q$};
		\node [style=none] (1) at (-.9, -0) {$1$};
		\node [style=none] (2) at (0.6, -0) {$=\frac{1}{p_0}$};
		\node [style=copoint] (3) at (4, 0.75) {$e_{\rho_0}$};
		\node [style=none] (4) at (2.5, -0) {};
		\node [style=none] (5) at (3.25, -0.5) {$\psi$};
		\node [style=none] (6) at (4, -0) {};
		\node [style=none] (7) at (4.75, -0) {};
		\node [style=none] (8) at (1.75, -0) {};
		\node [style=none] (9) at (3.25, -1) {};
		\node [style=none] (10) at (6, -0) {$:=$};
		\node [style=copoint] (11) at (7.5, 0.75) {$e_{\rho_0}$};
		\node [style=point] (12) at (7.5, -0.75) {$\rho_0$};
		\node [style=none] (13) at (10, -0) {$=\sum_\beta p_\beta$};
		\node [style=point] (14) at (12.5, -0.75) {$\beta$};
		\node [style=copoint] (15) at (12.5, 0.75) {$e_{\rho_0}$};
	\end{pgfonlayer}
	\begin{pgfonlayer}{edgelayer}
		\draw (8.center) to (7.center);
		\draw (7.center) to (9.center);
		\draw (9.center) to (8.center);
		\draw (0) to (4.center);
		\draw (3) to (6.center);
		\draw (11) to (12.center);
		\draw (15) to (14.center);
	\end{pgfonlayer}
\end{tikzpicture}
\]
Hence, from Eq.~(\ref{picked out with probability 1}) it follows that $e_{\rho_0}[\beta]=1$ where $e_{\rho_0}$ is the pure effect that steers to $s$ and $\beta$ a pure state. Transitivity then implies that for any pure state there is some pure effect which picks it out with probability $1$.

\paragraph*{Proof of lemma~\ref{pure state on max. mixed}:}
This result was proved in proposition $11$ of \cite{chiribella2015entanglement} and lemma $30$ of \cite{chiribella2011informational}. The conjunction of lemma~\ref{pure effect}, the generalised Choi theorem (see the proof of lemma~\ref{pure effect} for a brief discussion of this point), and steering (again see the proof of lemma~\ref{pure effect} for a brief discussion) is all that is needed for these proofs to go through, hence lemma~\ref{pure state on max. mixed} follows.
\endproof

\end{document}